%% file: MTT_ReEdit.tex
\documentclass[preprint,eqsecnum,showpacs,nofootinbib,
aps,amsmath,amssymb, tightenlines]{revtex4}

\usepackage{graphicx,color,mathrsfs}

\def\Lie{{\cal L}}
\def\l{\ell}

\def\thl{\theta_{(\l)}}
\def\thn{\theta_{(n)}}
\def\V{{\cal V}}
\def\e{\tilde{\epsilon}}

\def\ba{\begin{eqnarray}}
\def\ea{\end{eqnarray}}
\def\be{\begin{equation}}
\def\ee{\end{equation}}

\begin{document}


\begin{flushright}
IGPG-05/6-5\\
gr-qc/0506119
\end{flushright}


\title{Marginally trapped tubes and dynamical horizons}


\author{Ivan Booth}
\email{ibooth@math.mun.ca}
\affiliation{Department of Mathematics and Statistics,
Memorial University of Newfoundland, St.~John's, Newfoundland
and Labrador, A1C 5S7, Canada}

\author{Lionel Brits}
\email{lionelbrits@phas.ubc.ca}
\affiliation{Department of Physics and Physical Oceanography,
Memorial University of Newfoundland, St.~John's, Newfoundland
and Labrador, A1C 5S7, Canada}

\author{Jose A. Gonzalez}
\email{Jose.Gonzalez@uni-jena.de}
\affiliation{Theoretical Physics Institute, 
University of Jena, Max-Wien Platz 1, 07743 Jena, Germany}

\author{Chris Van Den Broeck}
\email{vdbroeck@gravity.psu.edu}
\affiliation{Institute for Gravitational Physics and Geometry, 
Department of Physics, The Pennsylvania State University, 
104 Davey Laboratory, University Park, PA 16802, USA}


\begin{abstract}
We investigate the generic behaviour of marginally trapped tubes 
(roughly time-evolved apparent horizons) using simple, spherically 
symmetric examples of dust and scalar field collapse/accretion onto 
pre-existing black holes. We find that given appropriate physical 
conditions the evolution of the marginally trapped tube may be either 
null, timelike, or spacelike and further that the marginally trapped 
two-sphere cross-sections may either expand or contract in area. 
Spacelike expansions occur when the matter falling into a black hole 
satisfies $\rho - P \leq 1/A$, where $A$ is the area of the horizon 
while $\rho$ and $P$ are respectively the density and pressure of the 
matter. Timelike evolutions occur when $(\rho - P)$ is greater than 
this cut-off and so would be expected to be more common for large 
black holes. Physically they correspond to horizon ``jumps" as 
extreme conditions force the formation of new horizons outside of 
the old. 
\end{abstract}



\maketitle


\section{Motivation}

Over the past decade, new definitions of black hole horizons have 
emerged which provide powerful tools for studying the behavior of 
black holes in the strongly dynamical regime. These ideas share 
the common philosophy that black holes should be thought of as 
physical objects in a spacetime that may be identified by local 
measurements. By contrast, traditional black holes and event horizons 
are globally defined properties of the causal structure of a 
spacetime \cite{hawkingellis72,wald}.

Though there has always been a certain amount of interest in the dynamics of
apparent horizons and their relation to black hole physics 
(see for example \cite{collins}), Hayward began this line of research in earnest with 
his definition of \textit{trapping horizons}. These were initially 
used to formulate dynamic versions of the laws of black hole mechanics in a 
quasi-local context  \cite{hayward94a}. Since then however, they have found a variety of applications
including, for example, studying interactions between black holes and gravitational waves
\cite{haywardPert1, haywardGravWav}.
A bit later, the \textit{isolated horizons} of  Ashtekar et 
al.~\cite{ashtekar,ashtekar02a} were developed to provide
a quasi-local characterization of the equilibrium states of black holes. In those works,
 it was shown that these objects obey a phase space formulation of the 
zeroth and first laws of black hole mechanics. Further, loop quantum gravity 
has made important use of them as boundary conditions in calculations of black hole
entropy \cite{entropy}.

Closely related to both trapping and isolated  horizons are \textit{dynamical horizons} \cite{ashtekar02b,ashtekar03a}, which characterize the dynamical phase of smooth black hole
evolutions. In particular, it has been shown that flux laws can be formulated for these horizons
that measure the growth of such quantities as energy and entropy. These 
laws contain terms that may be identified with fluxes of particular physical quantities 
such as matter and gravitational waves. A similar law exists for trapping horizons 
\cite{conservation} and it has been shown that the energy 
expressions involved agree with those derived in a Hamiltonian analysis of 
horizons as spacetime boundaries \cite{thebeast05}. 

In other developments, studies have been made of the perturbative, ``almost isolated"
regime \cite{haywardPert1,haywardPert2, booth04a} and 
isolated horizons have been given a convenient characterization in 
terms of multipole moments, which can be extended to a definition for 
the multipole moments for dynamical and trapping horizons 
\cite{multipoles}. Mathematical investigations have been made into 
such properties as existence and uniqueness \cite{abhaygreg,ams05} and recently
these notions have also begun to find application in the physical interpretation of numerical results \cite{num1,num2,num3}.

Thus, isolated, dynamical and trapping horizons constitute an increasingly
well-developed quasi-local framework for analytical as well as 
numerical studies of black hole dynamics in the strong field regime. 
However, in order to make effective use of them it is important to 
have clear intuition about how they behave. Unfortunately, there is a
paucity of good, analytical examples of spacetimes containing 
dynamical and/or trapping horizons and this has recently given rise 
to some confusion about the generic behavior of dynamical and 
trapping horizons. In this paper we will attempt to clarify matters 
by presenting several analytic and numerical examples of spacetimes
containing these horizons. 

Before considering these in more detail, let us recall the 
relevant definitions. 
From Hayward \cite{hayward94a} we have:
\\ \\
{\bf Definition 1.}
A \textit{trapping horizon} $H$ is a hypersurface in
a 4-dimensional spacetime that is foliated by 2-surfaces (which we 
will take to be of spherical topology) such that $\thl|_H= 0$, 
$\thn|_H \neq 0$ and $\Lie_n \thl|_H \neq 0$.
A trapping horizon is called \textit{outer} if $\Lie_n\thl|_H < 0$,
\textit{inner} if $\Lie_n\thl|_H > 0$, \textit{future} if 
$\thn|_H < 0$ and \textit{past} if $\thn|_H > 0$.
\\ \\
In this definition, and what follows, $\l^a$ and 
$n^a$ are respectively the future-directed 
outgoing and ingoing null normals to a leaf of the foliation while 
$\theta_{(\ell)}$ and $\thn$ are the expansion of the congruences of curves generated by 
those vector fields. Further, it is assumed that $n_a$ has been extended so that it is  
surface generating (ie.\ $n \wedge dn = 0$) in some neighbourhood of $H$. Then, from the definition, 
outer trapping horizons have trapped surfaces ``just inside" them while inner trapping 
horizons have trapped surfaces ``just outside". 

We will mainly be interested in a slight generalization of 
\textit{future} trapping horizons that was recently introduced by 
Ashtekar and Galloway \cite{abhaygreg}:
\\ \\
{\bf Definition 2.}
A \textit{marginally trapped tube (MTT)},  $T$ is a hypersurface
in a 4-dimensional spacetime that is foliated by two-surfaces 
(again assumed to be of spherical topology) such that 
$\thn|_T < 0$ and $\thl|_T = 0$.
\\ \\ 
We refer to the leaves of the foliation as 
\textit{marginally trapped surfaces}.\footnote{This follows the 
terminology used in \cite{abhaygreg}. Note however that the exact 
definition of marginally trapped surfaces varies somewhat in the 
literature. For example, Wald \cite{wald} defines a marginally 
trapped surface to be a compact, spacelike, two-surface for which 
both null-expansions are non-positive.}
MTTs have no restriction on their signature, which is allowed to
vary over the hypersurface. However, if an MTT is everywhere 
spacelike it is referred to as a \emph{dynamical horizon} 
\cite{ashtekar02a,ashtekar03a}, if it is everywhere timelike then 
it called a \emph{timelike membrane} (TLM) \cite{ashtekar03a,abhaygreg}, and 
if it is everywhere null and non-expanding then we have an \emph{isolated 
horizon}.\footnote{More precisely, this is a non-expanding horizon 
of which isolated horizons are a special case. However, we will 
follow the common usage of isolated horizon in its general rather 
than specific meaning throughout this paper.}

Note that the distinction between dynamical horizons and timelike 
membranes is more than just a technical difference of signature.  
For example, it is clear that since dynamical horizons are spacelike 
they may only be crossed in one direction by causal curves; 
this is a key characteristic of black hole horizons. By contrast, 
a timelike membrane, being timelike, obviously does not share this 
property. Further, it may be shown that while dynamical horizons always expand 
as they evolve, timelike membranes always shrink. Finally, in contrast to 
the dynamical horizon flux laws of \cite{ashtekar02a,ashtekar03a}, 
both the geometrical and matter contributions have indefinite 
signatures in the corresponding equations on timelike membranes. 
In particular, the geometric term no longer has a natural 
interpretation as a flux of gravitational radiation energy.  

Simple explicit examples of trapping horizons/MTTs are provided by the
Vaidya spacetimes \cite{ashtekar03a}. These are spherically symmetric
solutions to Einstein's equations describing the formation/growth of 
a black hole as null dust falls in from infinity. In the absence of a cosmological 
constant, the growth of the resulting black hole is always characterized by a 
dynamical/future outer trapping horizon. In the presence of a cosmological constant
the situation is a little more complicated as a second MTT appears --- a timelike membrane 
which can be associated with the cosmological horizon. Still, even in this case, the MTT associated
with the black hole remains a dynamical horizon. 

Heuristic arguments presented in \cite{ashtekar03a} and 
\cite{hayward00a} suggest that under physically reasonable
circumstances the MTTs associated with black hole formation and 
growth will always be spacelike and so future \emph{outer} trapping 
horizons are generic in this context. \emph{This intuition 
needs to be amended}. In Oppenheimer-Snyder spacetimes
 \cite{oppenheimer39}, which describe the gravitational collapse of 
spherical dust clouds, it has been shown 
that timelike membranes, rather than dynamical horizons, appear
during the formation of black holes. These spacetimes are constructed out of
a piece of the closed FRW universe (which forms a 
homogeneous and isotropic dust ball) surgically inserted into a Schwarzschild 
spacetime. 
As this ``star'' is allowed to collapse the 
horizon structure develops in the following way. At first there are
no horizons. Then, as the dust reaches a critical density an isolated 
horizon forms that cuts off the continuing collapse from the rest of 
the universe (in this case it is also an event horizon). 
Coincidentally, a timelike membrane appears and contracts
inside the isolated horizon until it reaches zero area. 

In this case, the timelike membrane is clearly associated with the 
formation of the black hole and cannot be dismissed as  
``cosmological''. One might argue that the OS spacetime is not very 
physical. Nevertheless, the question remains as to what is the 
generic behavior of MTTs during the formation or growth of a black 
hole. When are they spacelike and when are they timelike? 
Equivalently, when do they grow in the way that one would intuitively 
expect and when do they shrink? 

\begin{figure}
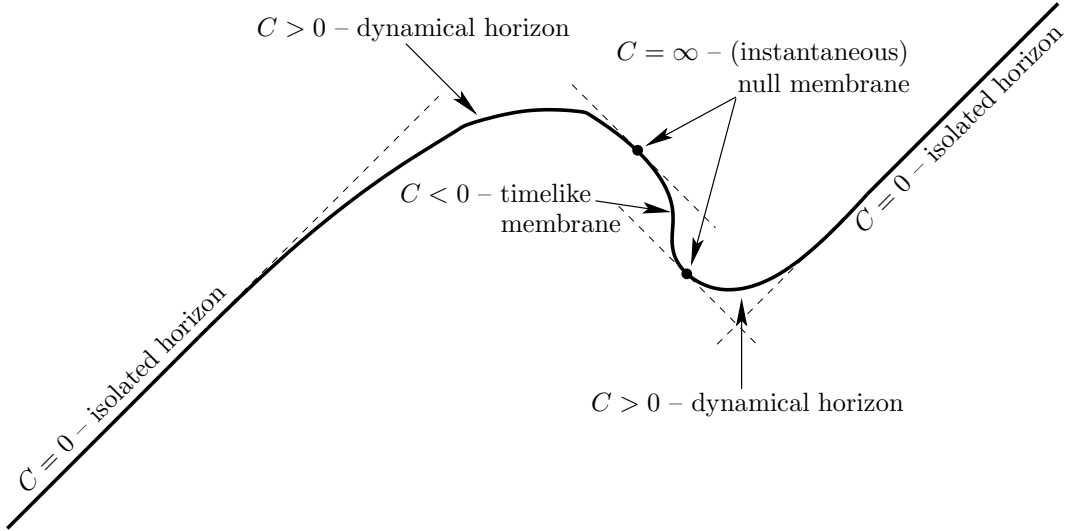
\caption{A schematic of an MTT as informed by our later examples. 
In this diagram $45^\circ$ lines are null, time increases in the 
vertical direction, while horizon area increases as one moves to 
the right.}
\label{MTTbehaviours}
\end{figure}

In this paper, we present several new examples of MTT-containing
spacetimes. From these, we see that in most circumstances
horizons are either isolated or dynamical, spacelike, and expanding. 
However, under certain conditions more interesting behaviours are 
seen. There are two ways in which new horizons may form where they 
did not exist before. 
In the first, an MTT may appear out of a singularity (such as a 
point of infinite density). In the second, large (though not 
infinite) concentrations of matter may force the creation of a 
dynamical horizon-timelike membrane pair. Such formations
are closely related to the well-known phenomena of 
``jumping" apparent horizons \cite{hawkingellis72}.

Horizons may also disappear.
Timelike membrane-dynamical horizon pairs can annihilate each other 
or alternatively MTTs can vanish into singularities. 
Figure \ref{MTTbehaviours} schematically displays some of these 
behaviours, and also suggests another interpretation. One can think 
of a single MTT that winds its way forwards and backwards through 
time rather than multiple dynamical horizons, timelike membranes, 
and isolated horizons that appear and disappear. 

For simplicity we will usually only consider spherically symmetric 
spacetimes.
Manifolds will be smooth unless stated otherwise. We use units such
that $G=c=1$, and spacetimes have signature $(-,+,+,+)$. We will 
assume that the Einstein equations hold, with matter satisfying the 
null energy condition.
 
The paper is structured as follows. In section \ref{mtt} we explain 
the basic properties of MTTs. In particular, we discuss how the 
signature of the metric induced on an MTT by the spacetime metric 
depends on the matter present and then examine several different 
types of matter. In section \ref{dust} we illustrate the analytic 
evolution of MTTs in various situations for the case of pressureless 
dust. In section \ref{scalarfields} we consider a couple of numerical 
examples of MTTs in the presence of a massive scalar field. 
Conclusions are presented in section \ref{conclusions}. 


\section{Marginally trapped tubes} 
\label{mtt}


\subsection{General properties}
\label{genprop}

As we have seen, the definition of MTTs is weaker than that of both 
dynamical horizons and timelike membranes (there is no restriction 
on the signature of the metric on $T$), and of trapping horizons
(there is no restriction that $\Lie_n \thl \neq 0$). On an MTT, both 
the induced metric signature and $\Lie_n \thl$ are allowed to vary. 

There are simple procedures which may be used to determine this signature. Here we will restrict
ourselves to the case of spherically symmetric spacetimes; 
the general case is closely related (see, for example, \cite{hayward94a}
or \cite{ashtekar03a}). With this condition, it is natural to restrict our attention to similarly symmetric MTTs.
\footnote{Non-spherically symmetric MTTs will also exist in these spacetimes. However, they are
more complicated to locate and so will not be considered in this first analysis. See
\cite{booth} and references therein for a discussion of what is know about the allowed behaviours
of such MTTs.} Further, it will be convenient to use the symmetry to foliate the spacetime 
into spacelike two-spheres and so extend the definitions of $\ell^a$, $n^a$ and $\thl$ off of $T$. 
We require that all such quantities share the symmetry and for simplicity will also impose the 
standard condition that $\ell \cdot n = -1$.

Next, following the conventions of 
\cite{booth04a, thebeast05}, we let $\V^a$ be a vector field 
which is: 1) tangential to $T$, 2) everywhere orthogonal to the foliation by 
marginally trapped surfaces, and 3) generates a flow which preserves 
the foliation. Thus, if $v$ is a foliation label, $\Lie_\V v$ is a 
function of $v$ only --- it is independent of the exact position on 
a leaf. Then it is always possible to find a function $C$ and normalization
of $\ell^a$  such that $\V^a = \l^a - C n^a$. Moreover, 
the definition of $\V^a$ implies that $\Lie_\V \thl = 0$, which gives 
us an expression for $C$:
\be
C = \frac{\Lie_\l \thl}{\Lie_n \thl} \, . 
\label{defC}
\ee
Note that $\V_a \V^a = 2  C$, so that the sign of $C$ determines the 
signature of $T$: if $C > 0$ it is spacelike, if $C = 0$ ($\Lie_\ell \thl = 0$)
or becomes undefined ($\Lie_\ell \thl \neq 0$  while $\Lie_n \thl = 0$) 
it is null, and if $C < 0$ it is timelike. In addition, the sign of 
$C$ determines whether $T$ is expanding, contracting, or unchanging 
in area.
Indeed, denoting the area element of the two-sphere cross-sections by
$\e$, one has
\be
\Lie_\V \e = - C \thn \e \, . \label{area}
\ee
This means that the expansion or 
contraction of an MTT  is linked to its signature (since $\thn < 0$).
In particular, when $T$ is spacelike it expands while when it is 
timelike it contracts. 

Further (\ref{defC}) explains the close relationship between the 
various trapping horizons and MTTs. If the null energy condition 
holds, then the numerator is non-positive (by the Raychaudhuri 
equation). Thus, the sign of $C$ is determined by the sign of 
$\Lie_n \thl$. Away from isolation and cases where 
$\Lie_n \theta_{(\ell)}=0$,\footnote{To understand why the latter implies
an important distinction, see the example in \cite{senovilla2}.} 
an MTT is a 
dynamical horizon if and only if it is a future outer trapping 
horizon. Similarly it is a timelike membrane if and only it is a 
future inner trapping horizon. 

To understand the generic behavior of these spherically symmetric 
MTTs we focus on this function $C$. 
Then, keeping in mind that $\thl=0$, from the Raychaudhuri equation 
it is easy to see that $\Lie_\l \thl = - 4 \pi T_{ab} \l^a \l^b$ while 
from $G_{ab} \ell^a n^b = 8 \pi T_{ab} \ell^a n^b$ one can show that 
$\Lie_n \thl = - \tilde{{\cal R}}/4 + 4 \pi T_{ab} \l^a n^b$, where 
$\tilde{{\cal R}}$ is the scalar curvature of the two-sphere 
cross-section of the MTT foliation (the easiest way to see this
is to consult a table of the Newman-Penrose equations, for example 
in \cite{stewart,chandra}). 
Then, using eq.~(\ref{defC}), 
\be
C = \frac{T_{ab} \l^a \l^b}{1/(2A) - T_{ab} \l^a n^b} \, , 
\label{F}
\ee
where $A$ is the area of a two-sphere cross-section of the MTT.
As noted above, if the null energy condition holds then the numerator 
is non-negative.
Thus, the sign of $C$ depends on the relative magnitude of 
$1/(2A)$ and $T_{ab} \l^a n^b$. 

The MTT behaviours implied by (\ref{defC}) and (\ref{F}) are closely related 
to Theorem 2 of \cite{ams05}, which may be summarized as follows. 
Suppose a (not necessarily spherically symmetric) 
spacetime is foliated by a family $\Sigma_t$ 
of spacelike hypersurfaces. Let $S \subset \Sigma_o$ be a marginally 
outer trapped surface, i.e., $\thl=0$ for an 
outgoing null normal $\l^a$ while the ingoing null expansion is unrestricted. 
Suppose $S$ is also ``strictly stably outermost'', which roughly means 
that if $S$ is deformed outward, the 
corresponding deformation of the outgoing null expansion is non-negative 
and positive somewhere. Then first of all, $S$ is contained in a horizon 
$H$ foliated by marginally outer trapped leaves that lie in $\Sigma_t$, which 
exists at least as long as these leaves remain strictly stably outermost. 
Moreover, if the null energy condition holds, $H$ is achronal. 
If $G_{ab} \l^a \l^b > 0$ somewhere on $S$, then $H$ is spacelike everywhere 
near $\Sigma_o$.

The condition that $S$ be strictly stably outermost is equivalent
to the requirement that a certain operator $L_{\Sigma_o}$ acting on 
functions $\psi$ on $S$ has a strictly positive principal eigenvalue. 
Restricting ourselves to spherical symmetry again and using the Einstein 
equations, this operator reduces to
\begin{equation}
L_{\Sigma_o} \psi = -\tilde{\Delta} \psi 
+ \left(\frac{1}{2}\tilde{\mathcal{R}} -8\pi T_{ab} \l^a n^b\right)\,\psi,
\end{equation}
where $\tilde{\Delta}$ is the Laplacian operator on the 
round 2-sphere $S$ and $\tilde{\mathcal{R}} = 8\pi/A$ the scalar 
curvature of $S$. The principal eigenvalue (corresponding to $\psi = \mbox{constant}$) is then
\begin{equation}
\lambda  = 8\pi \left[1/(2A) - T_{ab} \l^a n^b\right].
\end{equation}
According to the theorem in \cite{ams05}, if the null energy condition
holds then $\lambda > 0$ implies local achronality of the horizon, 
consistent with our considerations above. If, moreover, $G_{ab} \l^a \l^b 
= 8\pi T_{ab} \l^a \l^b > 0$ on $S$ then the horizon must be spacelike,
which is again as we found. Thus, the main results of \cite{ams05} applied
to the case of spherical symmetry are neatly encapsulated in the expression 
(\ref{F}) for the function $C$, which, however, will also tell us when we are 
dealing with a timelike membrane.

We now consider eq.~(\ref{F}) for some particular types of matter. 


\subsection{Behavior for some matter sources}


\subsubsection{Timelike perfect fluid} 
For a perfect fluid that moves along timelike worldlines with unit 
tangent $u^a$, the stress-energy tensor takes the form
\be
T_{ab} = (\rho + P) u_a u_b + P g_{ab} \, ,
\ee
where $\rho$ is the matter density of the fluid and $P$ is the 
pressure. Writing $u^a = \xi \l^a + (2\xi)^{-1} n^a$ for some 
function $\xi$, we find that
\be 
C = \frac{1}{2\xi^2}\frac{\rho+P}{(1/A) + P - \rho} \, , 
\label{perfectfluid}
\ee
and see that a priori, the MTT could have spacelike, null, or timelike evolution, the deciding factor being the magnitude of $(\rho - P)$ 
relative to $1/A$.

Simple examples of the various behaviours may be found in 
Robertson-Walker spacetimes \cite{bendov04a, senovilla1}. In the case where these 
cosmological models are collapsing, one can find spherical MTTs 
through all points in $M$. Picking a single MTT for definiteness 
(or equivalently selecting a ``centre" for the universe),
and assuming an equation of state of the form $P = \sigma \rho$, 
where $\sigma$ is some constant, these surfaces are:
i) timelike and contracting if $\sigma < 1/3$ (this includes a 
dust-filled universe, $\sigma=0$, and so the timelike contractions 
seen in Oppenheimer-Snyder collapse), ii) null and contracting if 
$\sigma = 1/3$ (a radiation filled universe with divergent $C$), and 
iii) spacelike and expanding if 
$\sigma > 1/3$.\footnote{Of course, these potential behaviours
are not confined to barotropic equations of state. In general,
given the dominant energy condition and assuming that $\rho \geq P$,
the signature of the MTT is determined by $\rho-3P$. } 
For now, we are not 
too concerned with the physical interpretation of such models and 
evolutions, but instead just note that these are all allowed 
mathematical possibilities. 

In the case of pressureless dust, $P=0$, one has the following 
heuristic picture deciding how the MTT will behave at a given 
two-sphere cross-section. Foliate spacetime by the level surfaces 
of dust particle proper time. Fix such a slice, let $r$ 
be some radial coordinate, and define the ``mass enclosed by a sphere 
with coordinate radius $r$'' to be $m(r) = m_o N(r)$, with $m_o$ the 
mass of a single particle and $N(r)$ the number of particles with 
radial coordinate smaller than $r$. Then eq.~(\ref{perfectfluid}) 
can be rewritten as
\be
C = \frac{f}{\tilde{\rho}_T - \tilde{\rho}}, 
\label{heuristic}  
\ee
where $f = R \rho/(4\xi^2) \geq 0$ with  $R = (A/4\pi)^{1/2}$ the areal 
radius, and $\tilde{\rho}_T$ and $\tilde{\rho}$ are two quantities 
with dimensions of surface density. $\tilde{\rho} = dm/dA$ is the 
change of enclosed mass with area, while $\tilde{\rho}_T$ is a purely 
geometric quantity associated with the MTT:
\be
\tilde{\rho}_T = \frac{1}{8\pi} M_T \tilde{{\cal R}},
\ee
with $M_T = R/2$ the instantaneous mass\footnote{See 
\cite{ashtekar03a} for a detailed discussion in the context of 
dynamical horizons; similar considerations hold for MTTs.}  of the MTT
and $\tilde{{\cal R}} = 8\pi/A$ the scalar curvature of the 2-sphere 
cross-section. $\tilde{\rho}_T$ is a straightforward generalization 
of the \emph{mass aspect} for isolated and dynamical horizons 
\cite{multipoles}: it determines the mass multipoles 
of the MTT and as such its definition and physical meaning are 
not restricted to spherical symmetry. Thus, the relative magnitude 
of the ``matter surface density'' $\tilde{\rho}$ and the ``geometric 
surface density'' $\tilde{\rho}_T$ is what governs the behavior of 
the MTT. 
Given a 2-sphere cross-section of the MTT, if the matter surface 
density is larger than the geometric surface density then the MTT 
will contract because of eq.~(\ref{area}), and in that case it must 
be timelike. If on the other hand the matter surface density is 
smaller than the geometric one, the MTT will expand, in which case 
it must be spacelike. When the two balance each other the MTT is null 
(though not isolated, as in this case $C \rightarrow \infty$ rather
than $0$). 


\subsubsection{Null fluids}
We next consider a null fluid that moves inwards from infinity 
towards some centre with tangent vector $n^a$. Then, the 
stress-energy tensor is
\be
T_{ab} = (\rho + P) n_a n_b \, ,
\label{nullfluid}
\ee
and so one quickly finds that 
\be
C = 2A\,(\rho + P) \geq 0 \, .
\ee
Hence, for matter of this type, the MTT can only be either isolated 
(if $\rho + P = 0$) or spacelike and expanding (otherwise). Indeed, 
this is the type of matter in the Vaidya spacetime where it is well 
known that MTTs demonstrate only these behaviours \cite{ashtekar03a}.

We note in passing that the Vaidya spacetimes can be generalized to
include both outgoing and ingoing null dust, plus a distribution of
energy whose rest frame is stationary. In such cases, the MTT
can again be timelike, as follows from the discussion in  \cite{israel1}. 
We suspect that with a careful choice of the parameters in these
models, one could also generate examples of MTTs that are partly 
timelike, partly null, and partly spacelike. However, we will not
discuss these examples further here, choosing instead to focus our 
attention on the more astrophysically relevant timelike dust spacetimes of 
section \ref{dust}.



\subsubsection{Scalar fields}
\label{scalar}
The final matter model that we will consider is that of a scalar 
field $\phi$. This has stress-energy tensor
\be
T_{ab} = \frac{1}{4\pi}\left(\nabla_a \phi \nabla_b \phi 
       - \frac{1}{2}\nabla_c \phi \nabla^c \phi\,g_{ab} 
       - V(\phi)\, g_{ab} \right) \, ,  
\label{scalarT}
\ee
where $V(\phi)$ is a potential function.  
For arbitrary potentials, $C$ takes the form
\be
C = \frac{2 (\Lie_\l \phi)^2}{(1/A) - 2 V(\phi)} \, . 
\label{Cscalar}
\ee 
If $V = 0$ (a massless scalar field), then
$\Lie_{\l} \phi \neq 0$ implies that the MTT must be spacelike. 
However, unless $V$ is negative definite, a non-zero 
potential a priori allows for spacelike, null, and timelike evolutions. 
In particular, this is the case for a massive Klein-Gordon field 
where $V(\phi) = m_o \phi^2/2$ for some $m_o > 0$.


\section{Pressureless dust}
\label{dust}

Examples of all of the MTT behaviours discussed above can be seen 
within the Tolman-Bondi family of solutions. In this section we 
will review this surprisingly rich set of solutions and then discuss 
specific members which display the various behaviours. 


\subsection{Tolman-Bondi solutions}
\label{TBsols}

These solutions describe the gravitational collapse of spherically 
symmetric dust clouds. They are very easy to work with and allow us 
to trace the evolution of a spacetime from specified initial
conditions. A nice discussion can be found in the early part of 
\cite{gonc} and with minor changes, we will follow that 
description below. 

Initial conditions are given on a spherically symmetric, 
spacelike, three-surface $\Sigma_o$. On that surface, we may specify: 
i) the dust density $\rho_o$, so that on that surface 
$T_{ab} = \rho_o u_a u_b$ where $u^a$ is the forward-in-time pointing 
timelike unit normal to $\Sigma_o$, and ii) the initial areal velocity
$v_o = \frac{dr}{d\tau}$ of the dust, where $r = \sqrt{A/4\pi}$ is the
areal radius and $\tau$ is the proper time as measured by observers 
comoving with the dust. Then, taking the areal radius $r$ as a 
coordinate on $\Sigma$ along with the usual spherical coordinates 
$(\theta, \phi)$, these two functions are sufficient to specify both 
the intrinsic metric $h_{ab}$ and extrinsic curvature $K_{ab}$ of 
$\Sigma_o$. Defining 
\ba
m(r) &=& 4 \pi \int_0^r \rho_o(\tilde{r}) \tilde{r}^2 
d\tilde{r} \label{mass}  \, \, \,  \mbox{and} \\
k(r) &=& \frac{2m(r)}{r} - v_o^2(r) \, , 
\ea
we have
\be
ds^2 = \frac{dr^2}{1 - k(r)} + r^2 (d \theta^2 
+ \sin^2 \theta d \phi^2) \, , 
\label{threemetric}
\ee
and
\be
K_{ab} = \frac{d v_o}{dr} \hat{r}_a \hat{r}_b 
+ \frac{v_o}{r} \Omega_{ab} \, , \label{TBK}
\ee
where $\Omega_{ab} = [d \theta]_a [d \theta]_b 
+ \sin^2 \theta [d \phi]_a [d \phi]_b$ and $\hat{r}^a$ is the 
spacelike, unit normal, outward pointing radial vector. With the 
restriction that $\Sigma_o$ be spacelike, we note that initial
conditions must be chosen so that $k(r)< 1$. 

The functions $m(r)$ and $k(r)$ have well-defined physical 
interpretations:  $m(r)$ is a mass function which measures the amount 
of matter contained within a sphere of areal radius $r$ while 
$k(r)$ determines whether or not the  system is gravitationally 
bound. We restrict our attention to gravitationally bound systems 
 ($k(r) > 0$) and so the allowed values of $k(r)$ are 
$0 < k(r) < 1$. Note too that if $v_o(r) = 0$, $\Sigma_o$ will represent an 
instant of time symmetry with $K_{ab} = 0$. Then 
$k(r) = \frac{2 m(r)}{r}$ and so $0 \leq 2 m(r) < r$. 

If we further restrict our attention to initial conditions for which all 
matter is initially either stationary or infalling ($v_o(r) \leq 0$), 
it is not hard to use the Einstein equations to show that with $\tau$ 
as the proper time measured by observers comoving with the dust, 
\be
\dot{R}(\tau,r) \equiv \frac{dR(\tau,r)}{d\tau} = 
- \sqrt{ \frac{2m(r)}{R(\tau,r)} - k(r) }\, , 
\label{EErem}
\ee
where $R(\tau,r)$ is defined as the areal radius at time $\tau$ of the dust 
shell that had initial areal radius $r$ on $\Sigma_o$.

Then, in Gaussian normal coordinates, these initial conditions 
can be evolved to give us a four-dimensional metric:
\be
ds^2 = - d\tau^2 + \frac{(R'(\tau,r))^2}{1 - k(r)}dr^2 
+ R^2(\tau,r)\, d \Omega^2 \, , 
\ee
where $R' = \partial R/\partial r$ and the stress-energy tensor takes 
the form $T_{ab} = \rho(\tau,r) \nabla_a \tau \nabla_b \tau$ with:
\be
\rho(\tau,r) = \frac{1}{4 \pi R^2(\tau,r)} \frac{dm}{dR} = 
\frac{r^2 \rho_o(r) }{R^2(\tau,r) R'(\tau,r)} \, .
\label{rho}
\ee

Now, there is an exact, parametric, solution to equation (\ref{EErem}). Specifically, for $0 \leq \eta < \pi$ 
and initial areal radius $r$:
\begin{eqnarray}
\tau(\eta,r) &=& \tau_o(r) + \frac{m(r)}{k^{3/2}(r)} (\eta + 
\sin \eta) \, \mbox{ and} \label{taueta} \\
R(\eta,r) &=& \frac{2m(r)}{k(r)} \cos^2 \left( \frac{\eta}{2} 
\right)\, , 
\label{Reta}
\end{eqnarray}
where
\be
\tau_o(r)=\frac{ r v_o(r)}{2 k(r)} + \frac{m(r)}{k^{3/2}(r)} 
\arccos \sqrt{1-\frac{r v_o^2(r)}{2m(r)}} \, .
\ee
In the special case where $v_o(r) = 0$, then $\tau_o(r) = 0$ and 
$\tau = 0$ corresponds to $\eta = 0$. 
For simplicity, our explicit examples will be restricted to 
such evolutions; it turns out that these are sufficient to demonstrate all of 
the potential MTT evolutions. The equations
(\ref{taueta}) and (\ref{Reta}) then reduce to:
\begin{eqnarray}
\tau(\eta,r) &=& \left(\frac{r^3}{8m(r)}\right)^{1/2}(\eta 
+ \sin \eta),\\ 
R(\eta,r) &=& r \cos^2\left(\frac{\eta}{2}\right).
\end{eqnarray}

In constructing our examples, we will sometimes find it convenient 
to excise the interior or exterior part of a dust spacetime and 
replace it by a Schwarzschild region. Then, the Einstein equations 
are satisfied at the three-dimensional junction surface if and only 
if its intrinsic and extrinsic curvatures are the same whether 
measured on the interior or the exterior of the surface. For our 
purposes, it will be sufficient to only consider excisions along 
the comoving surfaces of constant $r$. Specifically, suppose we 
wish to take the junction surface as $r = \hat{r}$, making the 
spacetime Schwarzschild either for $r<\hat{r}$ or $r>\hat{r}$. 
Then assuming $v_o=0$,  
the induced metric from the dust side on the corresponding 
3-dimensional timelike junction surface is
\begin{equation}
d\hat{s}^2 = -\frac{\hat{r}^3}{8m(\hat{r})}(1 + \cos\eta)^2 d\eta^2 
+ \hat{r}^2 \cos^4\left(\frac{\eta}{2}\right)\,(d\theta^2 
+ \sin^2\theta\,d\phi^2).
\end{equation}
The induced metric from the Schwarzschild side is the same with 
$m(\hat{r})$ replaced by the Schwarzschild mass $M$. The non-zero 
components of the extrinsic curvature from the dust side are
\begin{equation}
K_{\theta\theta} = \frac{K_{\phi\phi}}{\sin^2\theta} 
= -\hat{r}\cos^2\left(\frac{\eta}{2}\right)\sqrt{1
-\frac{2m(\hat{r})}{\hat{r}}},
\end{equation}
and from the Schwarzschild side we again get the same with 
$m(\hat{r})$ replaced by $M$. Thus the resulting restrictions are 
quite simple.
If it is the \emph{exterior} that is being replaced by Schwarzschild beyond some coordinate radius $\hat{r}$, then the matching of intrinsic 
and extrinsic curvatures of the boundary requires that the 
Schwarzschild mass of the exterior geometry be $M = m(\hat{r})$. 
If the \emph{interior} is being replaced by a Schwarzschild geometry 
with mass $M$ for $r < \hat{r}$, then the relationship between mass 
function and initial density should be
\begin{equation}
m(r) = M + 4\pi \int_{\hat{r}}^r \bar{r}^2 \rho_o(\bar{r}) d\bar{r}
\end{equation}
for $r \geq \hat{r}$.

Thus, given initial conditions,  we can analytically calculate the 
full, four-dimensional metric as long as the Gaussian normal 
coordinate system remains valid. As can be inferred from the 
preceding discussion, the reason for this simplicity is that  
any shell of radius $r$ effectively evolves along the geodesics of 
a Schwarzschild solution of mass $m(r)$ --- outer shells
do not affect the evolution of inner shells and inner shells only 
affect outer shells through the total mass function $m(r)$. 
This makes these spacetimes especially easy to work with but it 
also points to a potential problem. If the shells of constant $r$ 
do not maintain their original ordering, then the mass contained 
within a shell can change with time, and in this case the evolution 
will no longer be described by the solution discussed above. Further, 
apart from this physical problem, shell-crossings will also cause 
the Gaussian normal coordinates to break down as those surfaces of 
constant $r$ were also used as coordinates.

There is a large body of literature on shell crossings (see for 
example the references in \cite{gonc}), and the Tolman-Bondi 
spacetimes are often used to study this phenomenon. In this paper, 
however,  we are only interested in using these spacetimes to provide 
concrete examples of MTT evolutions. As such we will studiously avoid 
such complications. A sufficient condition to guarantee that no 
shell-crossings occur is easily seen to be 
\be
R'(\tau,r) \geq 0 \, , \label{shellCon}
\ee
as this will ensure a physical separation of $r = \mbox{constant}$ 
surfaces. 
If $v_o(r) = 0$ it is not hard to see that this reduces to 
\be
\frac{d \tau_c}{dr} \geq 0 \, , 
\label{taucon}
\ee 
where 
\be
\tau_c(r) =  \pi \sqrt{\frac{r^3}{8 m(r)}} \, , 
\label{tauc}
\ee
is the time for the shell of initial areal radius $r$ to collapse to 
zero area. 


\subsection{MTTs in Tolman-Bondi spacetimes}

The above considerations define the Tolman-Bondi spacetimes. We now 
consider the location of \emph{spherically symmetric} marginally trapped 
surfaces within those spacetimes.
For definiteness we choose
\be
\ell_a = - [d \tau]_a + \frac{R'(\tau,r)}{\sqrt{1-k(r)}} [dr]_a 
\, \, \,  \mbox{ and} \, \, \,
n_a =  - \frac{1}{2} [d \tau]_a - \frac{R'(\tau,r)}{2 \sqrt{1-k(r)}} 
[dr]_a \, , 
\ee
as the forward-in-time outward and inward null vectors 
satisfying $\ell \cdot n = -1$. Note too that these expressions 
will always be well-defined since our requirement that $\Sigma_o$ be 
purely spacelike fixes $k(r) < 1$.

Of course $\l$ and $n$ can be rescaled, but for the purposes of this 
paper that scaling is irrelevant (we only care about whether 
quantities defined with respect to them are zero, negative or 
positive) and so we can choose to work with this convenient choice of 
vectors. Then, keeping in mind that $\thl = (g^{ab} + \ell^a n^b 
+ n^a \ell^b) \nabla_a \ell_b$ (and similarly for $\thn$):
\be
\thl = \frac{2 ( \dot{R}(\tau,r) + \sqrt{1-k(r)} )}{R(\tau,r)}  
\, \, \, \mbox{and} \, \, \, 
\thn = \frac{\dot{R}(\tau,r) - \sqrt{1- k(r)} }{R(\tau,r)} \, .
\ee
Solving $\thl=0$ with a bit of help from evolution equation 
(\ref{EErem}), it is not hard to see that marginally trapped surfaces 
occur whenever
\be
R(\tau,r) = 2 m(r) \, . \label{Rloc}
\ee
This is the expected result if one recalls that shells of constant 
$r$ essentially move along the geodesics of a Schwarzschild spacetime 
with mass $m(r)$. Furthermore, it is clear that on such a surface, 
\be
\thn =  -\frac{ \sqrt{1- k(r)} }{ m(r)}  < 0 \, .
\ee
Thus, in these spacetimes, any two-sphere on which $\thl=0$ is part 
of an MTT.  

Finally we can calculate the expansion parameter $C$. Using condition 
(\ref{Rloc}), definition (\ref{defC}), the evolution equation
(\ref{EErem}), and definition (\ref{mass}) one can directly calculate
\be
C = \frac{2m'(r)}{R'(\tau,r)-m'(r)}  = \frac{2 \rho(\tau,r)}{1/A
- \rho(\tau,r)} \, , \label{dustC}
\ee
where $A = 16 \pi m^2(r)$. As would be expected this result agrees 
with the earlier, more general, perfect fluid results 
(\ref{perfectfluid}) and  (\ref{heuristic}). 

With this background established, generating examples of MTT 
spacetimes is simply a matter of picking initial conditions and 
using (\ref{taueta}) and (\ref{Reta}) to generate the evolution. The 
expression for $C$, eq.~(\ref{dustC}) can then be used to determine 
the signature of the tube. Following these simple procedures we 
generate the examples below, all of which will start from an instant 
of time symmetry ($v_o = 0$). Often we will consider situations where 
the dust accretes onto a pre-existing hole. In those cases, excisions 
will be performed inside of some shell so that a black hole may be 
inserted into the spacetime. When this is done we will use the 
Schwarzschild mass  of the interior as a reference scale for 
masses, lengths, and times. In other cases the scale will be set 
according to the physics of the particular situation.


\subsection{Dust ball collapse}
\label{dustcollapse_sec}


\subsubsection{Collapse of a dust ball with Gaussian initial density}

\begin{figure}
\begin{minipage}{6cm}
\resizebox{6cm}{6cm}{\includegraphics{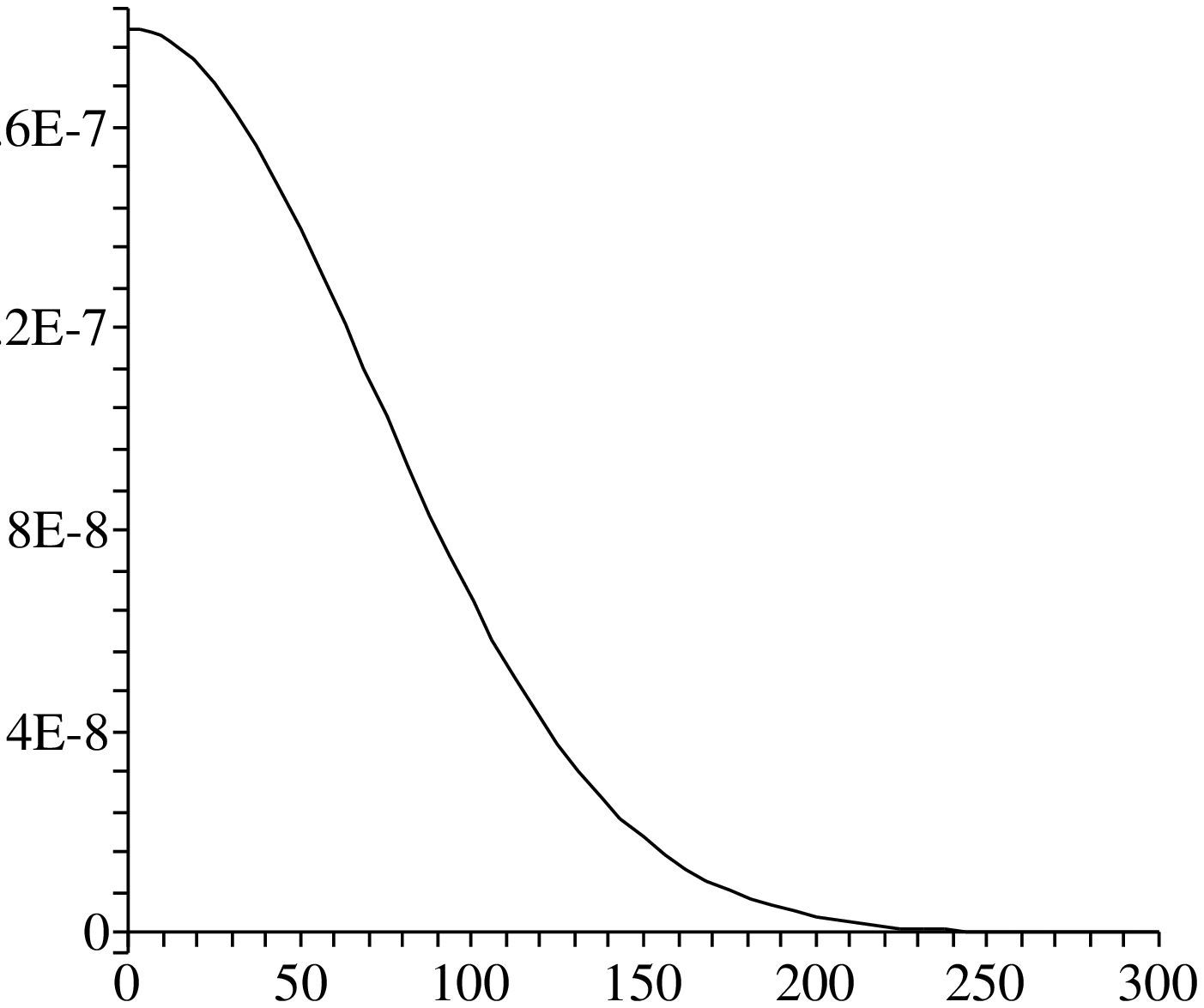}}
\put(-110,0){a) $\rho_o$ vs $r$}
\end{minipage}
\begin{minipage}{6cm}
\resizebox{6cm}{6cm}{\includegraphics{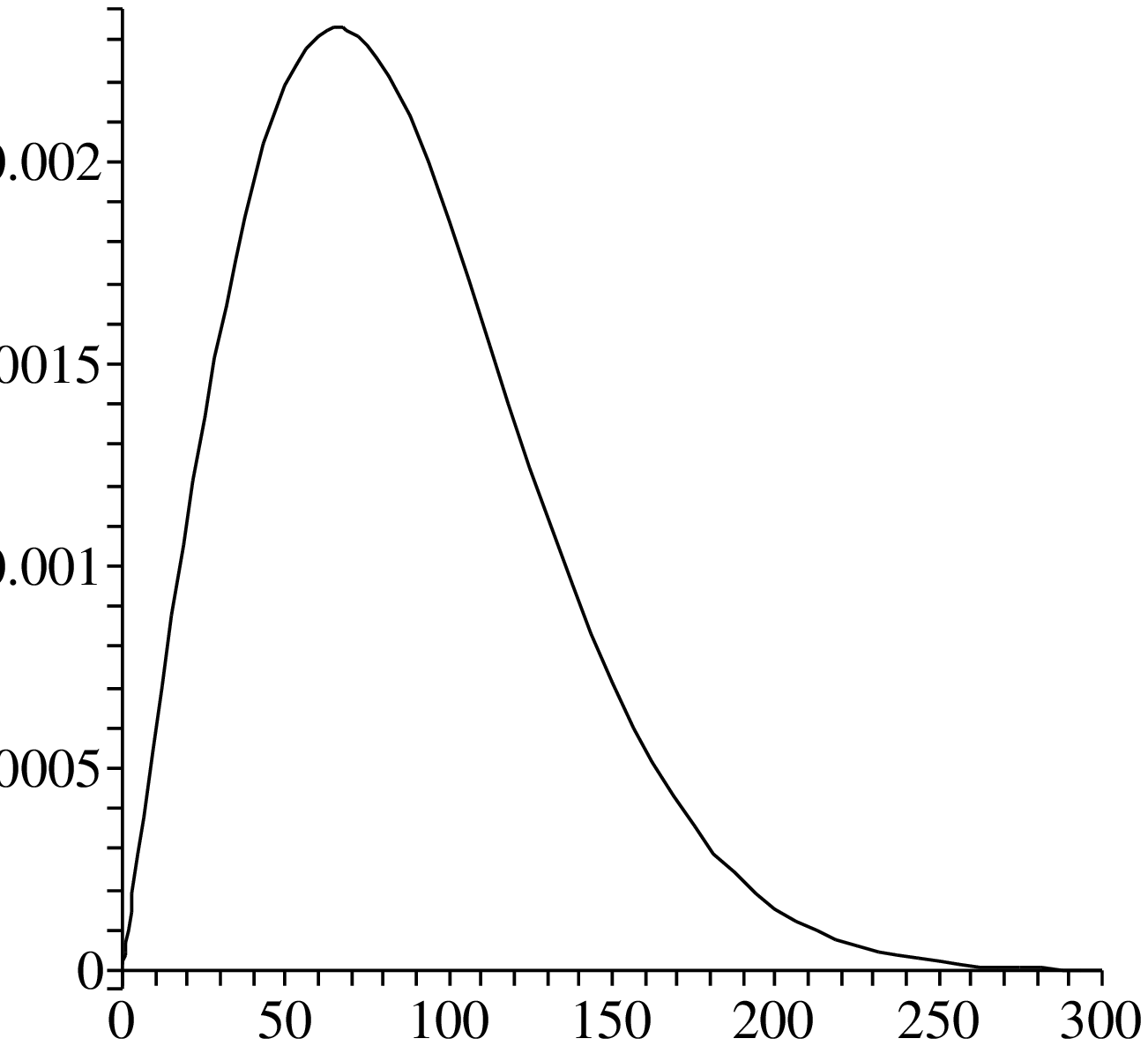}}
\put(-110,0){b) $C$ vs $r$}
\end{minipage}
\begin{minipage}{8cm}
\resizebox{8cm}{8cm}{\includegraphics{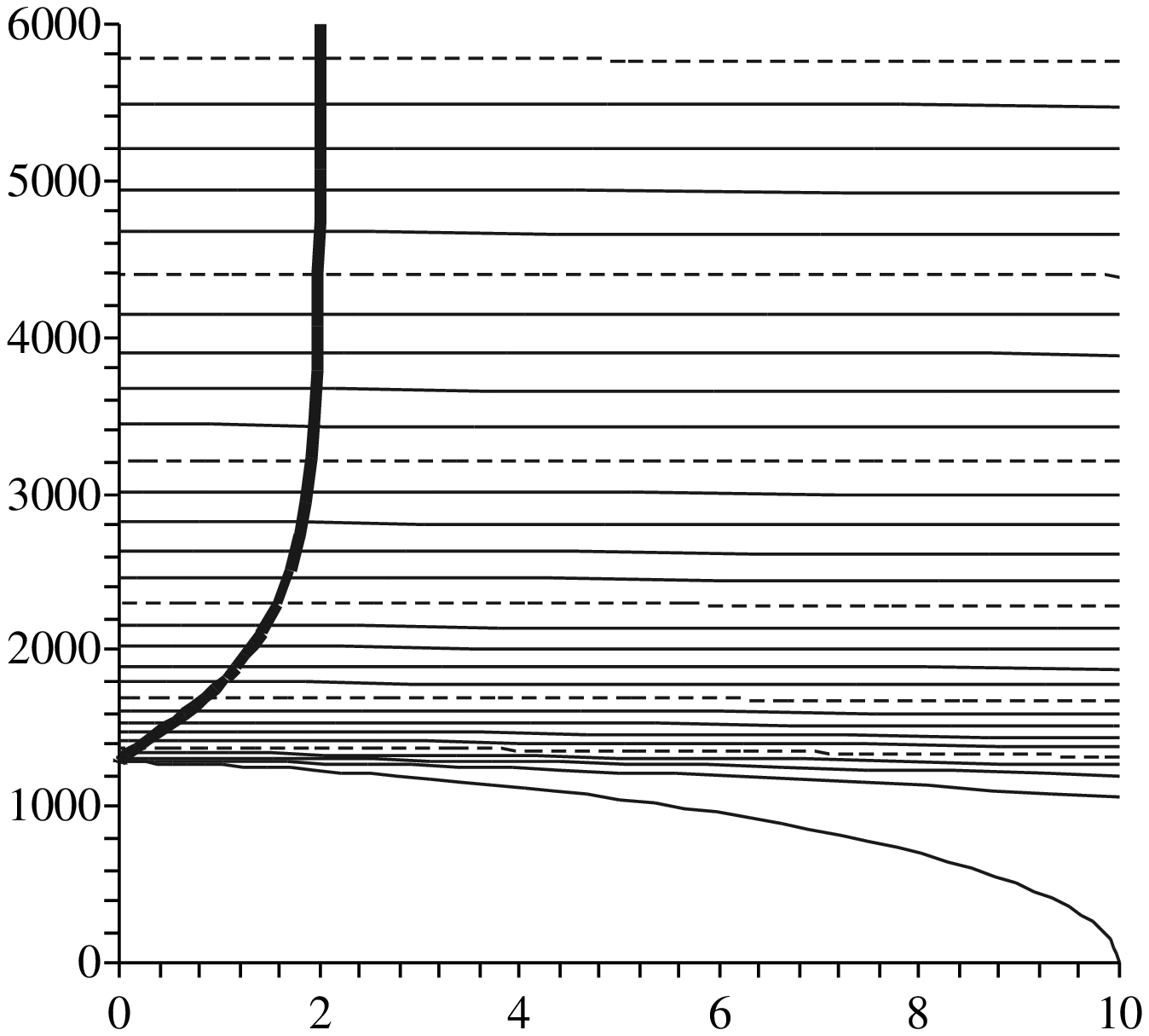}}
\put(-170,0){c) $\tau$ vs $R$ for MTT evolution}
\put(-25,60){ {\footnotesize{$r=50$}}}
\put(-25,70){ {\footnotesize{$r=100$}}}
\put(-25,90){ {\footnotesize{$r=150$}}}
\put(-25,118){ {\footnotesize{$r=200$}}}
\put(-25,154){ {\footnotesize{$r=250$}}}
\put(-25,196){ {\footnotesize{$r=300$}}}
\end{minipage}
\caption{Collapse of a dust ball}
\label{dustcollapse}
\end{figure}

We begin with the example of a non-uniform dust ball collapsing to 
form a black hole. The initial density distribution is taken to be 
radially Gaussian so that
\be
\rho_o = \frac{m_o}{\pi^{3/2} r_o^3} e^{-r^2/r_o^2}\, .
\ee 
In this expression $m_o$ is the total mass of the cloud as read 
from the corresponding mass function $m(r)$ (equation \ref{mass}) and
also corresponds to its ADM mass. 
The parameter $r_o$ determines how much the cloud is initially 
``spread out". Such a distribution is pictured in figure 
\ref{dustcollapse}a{\footnotesize)}, where we have taken 
$r_o = 100 \,m_o$. $\rho$ is plotted in units of $m_o^{-2}$ and 
$r$ in units of $m_o$.

For this choice of parameters $r - 2 m(r) > 0$ everywhere and so 
there are no marginally trapped surfaces in the initial time slice. 
Further, for any distribution of this 
kind condition (\ref{taucon}) is met and therefore there are no shell 
crossings. 
In this case, 
\be
\tau_c(0) = \frac{ \pi^{5/4} \sqrt{3}}{4 \sqrt{2}} 
\sqrt{\frac{r_o^3}{m_o}} \approx 1281\,m_o
\, ,
\ee
and the first $r = \mbox{constant}$ shells collapse to zero 
area at that time. As this happens, the central density 
$\rho(\tau_c, 0)$ diverges to infinity (equation \ref{rho}). 
Referring to \ref{dustcollapse}c{\footnotesize)} we see that an MTT 
is born out of this divergence, while 
\ref{dustcollapse}b{\footnotesize)} shows that it is everywhere 
spacelike and so is a dynamical horizon. 

Note that in figure \ref{dustcollapse}c{\footnotesize)} (as well as 
in future evolution graphs), the MTT is shown as the thick black line. 
The rest of the lines are $r = \mbox{constant}$ surfaces and 
correspond to the timelike geodesics that trace the evolution of 
individual dust shells. 
In this figure, most of them are very nearly horizontal as they have 
fallen in from a long way out and are moving very quickly (relative 
to the constant $\tau$ foliation) by the time that they approach the 
horizon. 

Returning to the evolution graph \ref{dustcollapse}c{\footnotesize)} 
we note that as $\tau \rightarrow \infty$, $R \rightarrow 2 m_o$ and 
$C \rightarrow 0$ as the last bits of matter fall through the horizon 
and it asymptotes towards a null and isolated state.

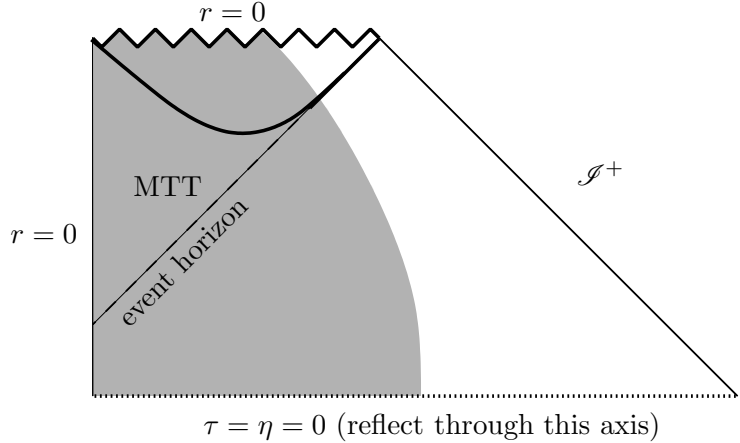
\begin{figure}
\input{SpacetimeCollapse.pstex_t}
\caption{Penrose-Carter diagram for dust-ball collapse. To obtain the full spacetime diagram, including 
the corresponding white hole and $\mathscr{I}^-$, reflect this diagram through the $\tau = \eta = 0$ instant of time symmetry. Note that $r=0$ starts as a regular timelike geodesic in the centre of the ball and then changes into a spacelike singularity at $\tau = \tau_c (0)$ as the central density goes to infinity.}
\label{STC}
\end{figure}

Many of these features of the evolution may also be seen in the Penrose-Carter diagram
for the spacetime which is given in figure \ref{STC}. Thus, we again see the MTT created out of
the central spacelike singularity, evolving in a spacelike fashion, and finishing its evolution by asymptoting
to the null event horizon. Note that for simplicity of presentation, this diagram shows a sharp cut-off of the dust distribution 
at some finite $r$. In our example this does not occur and instead the density asymptotes to zero. Thus, properly one should view the cut-off as as marking, say, the $r$ which contains $99.9 \%$ of the mass. This diagram also nicely
shows that while the MTT appears at the same time as the singularity, nothing in particular is happening at $r=0$ when the event horizon ``appears".


\subsubsection{Smooth versions of Oppenheimer-Snyder collapse}
\label{osexamples}

\begin{figure}
\begin{minipage}{6cm}
\resizebox{6cm}{6cm}{\includegraphics{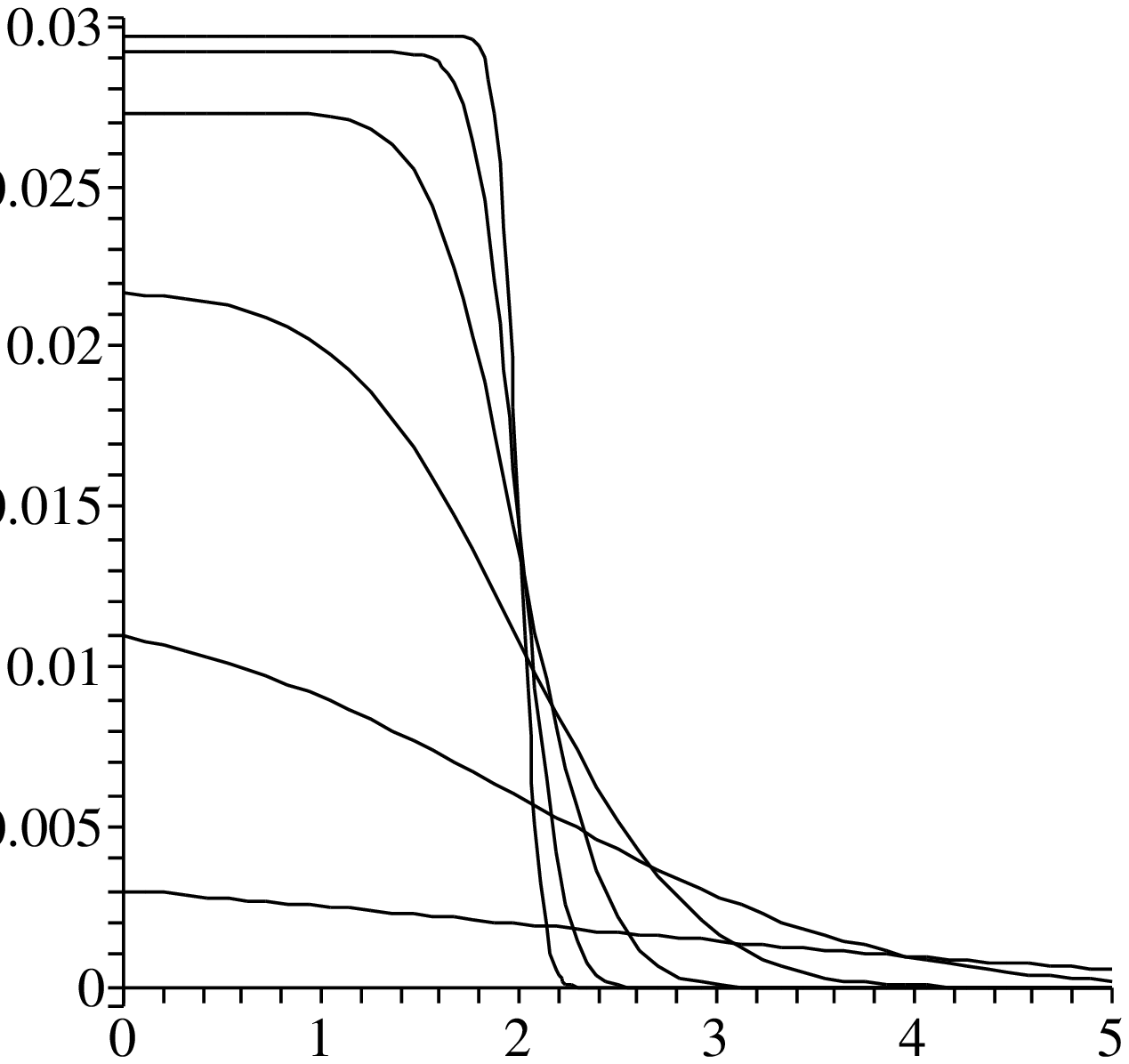}}
\put(-110,0){a) $\rho_o$ vs $r$}
\put(-155,30){ {\footnotesize{$\sigma=\frac{1}{2}$}}}
\put(-155,60){ {\footnotesize{$\sigma=1$}}}
\put(-155,100){ {\footnotesize{$\sigma=2$}}}
\put(-155,132){ {\footnotesize{$\sigma=4$}}}
\put(-133,142){ {\footnotesize{$\sigma=8$}}}
\put(-105,150){ {\footnotesize{$\sigma=16$}}}
\end{minipage}
\begin{minipage}{6cm}
\resizebox{6cm}{6cm}{\includegraphics{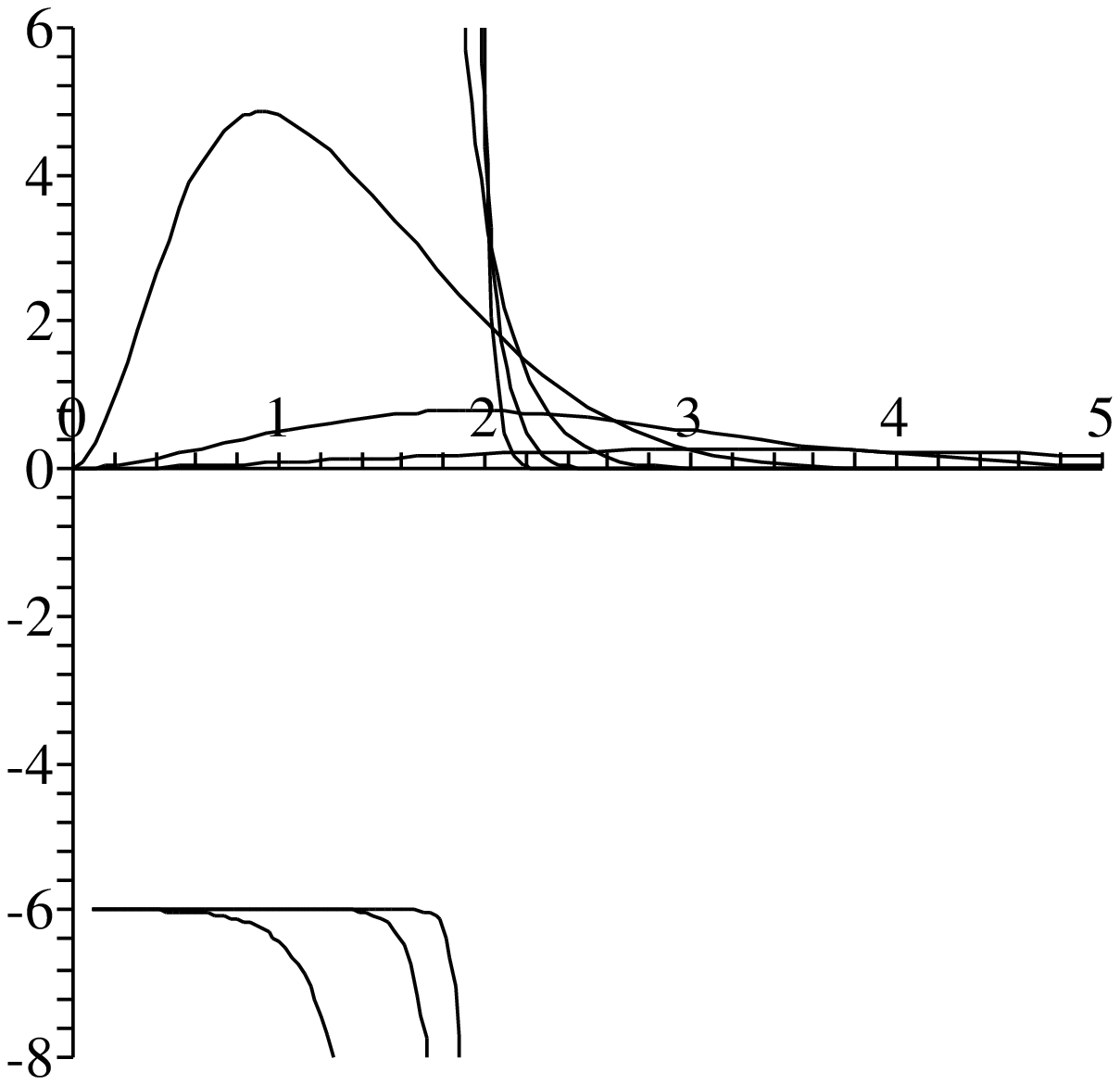}}
\put(-100,0){b) $C$ vs $r$}
\put(-130,83){ {\footnotesize{$\sigma=\frac{1}{2}$}}}
\put(-129,104){ {\footnotesize{$\sigma=1$}}}
\put(-143,147){ {\footnotesize{$\sigma=2$}}}
\put(-155,25){ {\footnotesize{$\sigma=4$}}}
\put(-134,10){ {\footnotesize{$\sigma=8$}}}
\put(-103,25){ {\footnotesize{$\sigma=16$}}}
\end{minipage}
\begin{minipage}{8cm}
\resizebox{8cm}{8cm}{\includegraphics{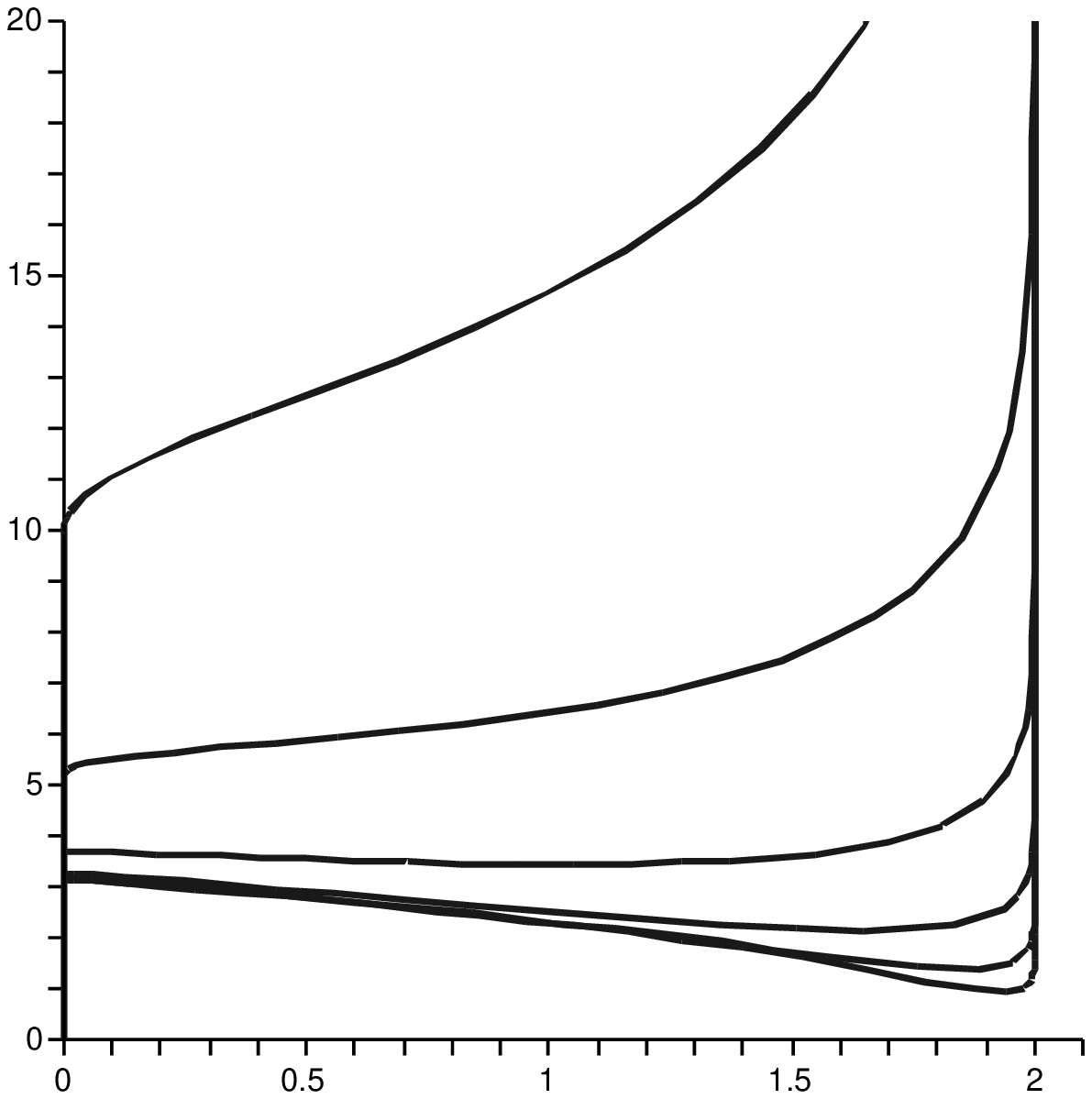}}
\put(-170,0){c) $\tau$ vs $R$ for MTT evolution}
\put(-50,25){ {\footnotesize{$\sigma=16$}}}
\put(-62,37){ {\footnotesize{$\sigma=8$}}}
\put(-100,45.5){ {\footnotesize{$\sigma=4$}}}
\put(-150,58){ {\footnotesize{$\sigma=2$}}}
\put(-150,83){ {\footnotesize{$\sigma=1$}}}
\put(-150,160){ {\footnotesize{$\sigma=\frac{1}{2}$}}}
\end{minipage}
\caption{Approach to Oppenheimer-Snyder collapse}
\label{OS}
\end{figure}

We now consider a family of spacetimes parametrized by a real 
number $\sigma$ where the initial density profiles are also not 
homogeneous but uniformly converge to a step function as 
$\sigma \rightarrow \infty$. These may then be viewed as 
interpolating between the collapse of a highly non-homogeneous 
dust ball like in the previous example, and Oppenheimer-Snyder
collapse. As one would expect, in these examples we will encounter 
marginally trapped tubes that are partially spacelike and partially 
timelike.

In particular, consider initial densities of the 
form\footnote{Similar initial density profiles can more easily
be constructed by means of $\tanh$ functions. However, the associated
mass functions $m(r)$ are awkward expressions in terms of polylogarithms,
which are multi-branched and need to be treated with care. By defining 
initial density distributions in terms of error functions we avoid such
complications.}
\begin{equation}
\rho_o =  \frac{m_o F(\sigma)}{r_o^3} \left( 1 
- \mbox{erf} \left[ \sigma \left(\frac{r}{r_o} 
- 1\right) \right] \right)
\label{OSlikedensity}
\end{equation}
where $F(\sigma)$ is a complicated function\footnote{For those who 
are interested: $\displaystyle F (\sigma) = 
\frac{3\sigma^3}{2\pi \sigma ( 2 \sigma^2 + 3)(1 
+ \mbox{erf} \, \sigma) + 4 \sqrt{\pi} e^{-\sigma^2} ( 1 
+ \sigma^2).}$}
chosen so that $m_o = \lim_{r \rightarrow \infty} m(r)$ is the total 
dust mass, $r_o$ is the location on the ``step" where $-d \rho / d r$ 
is a maximum, $\sigma$ characterizes the steepness of that step, and 
$\mbox{erf}(x)$ is the usual error function.


In the examples of Figure \ref{OS} we take $m_o$ as our length 
scale, choose $r_o = 2 m_o$, and consider a variety of values of 
$\sigma$. Then graph a) shows how the initial densities converge 
towards a step function with increasing $\sigma$. In the meantime 
graph b) shows that for smaller values of $\sigma$ the expansion 
parameter $C$ is always positive, while for the larger values it 
starts out negative with $C\approx - 6$ while the density is 
approximately constant. This corresponds to the expected value of 
$C$ in the corresponding OS spacetime. It then diverges to $- \infty$,
switches to being positive around $r = 2m_o$ (when the main part of the step
has fallen through), and then asymptotes to $0$ as the density of dust falling 
through the horizon also goes to zero. As expected, the switch in 
sign corresponds to the corresponding switch of $\Lie_n \thl$ --- 
that is when the MTT  becomes instantaneously tangent to $n^a$. 

In c) we also see that in all cases a dynamical horizon asymptotes 
to $R = 2 m_o$ as the last bits of dust fall through it. However, 
the MTT behaviour before that time varies greatly. For small values 
of $\sigma$ it continuously increases in area. For large values, we 
see that there are both increasing and decreasing regions. Somewhat 
confusingly it appears that for $\sigma = 2$ there are both 
increasing and decreasing regions as well, even though b) shows that 
$C > 0$ everywhere. We will return to examples such as this in 
section \ref{semh}, but for now we just keep in mind that spacelike 
surfaces can intersect in non-trivial ways. Here the dynamical 
horizon is demonstrating this as it intersects some of the 
$\tau = \mbox{constant}$ surfaces twice. The possibility of
such foliation effects was also noted in \cite{ams05}.

Note that the timelike membranes in this example all go to 
zero areal radius around $\tau = \pi m_o$. This is not surprising as 
for large values of $\sigma$, $\tau_c \approx \pi m_o$ (eq. 
(\ref{tauc})). Thus, the membranes vanish as the first dust shells
also collapse to zero areal radius --- that is they disappear into 
the density singularity. 

Finally, note that the spacetime diagrams for these spacetimes would be very similar to 
that shown in figure \ref{STC}. The only significant difference would be that for the spacetimes
with timelike membranes, the MTT would emerge from/vanish into 
the singularity as a timelike rather than spacelike surface -- the slope of the MTT would be greater
than $45^\circ$.

%






\subsection{Accretion onto a pre-existing hole}
\label{accrete}

The next two examples study the accretion of dust shells onto a 
pre-existing black hole. In the first example a black hole will 
substantially increase its mass while the MTT remains null or 
spacelike everywhere through the evolution. In the second example 
we study a very large matter shell falling into a black hole; here 
we will again see timelike membranes.


\subsubsection{Small dust shell falls into black hole}

\begin{figure}
\begin{minipage}{6cm}
\resizebox{6cm}{6cm}{\includegraphics{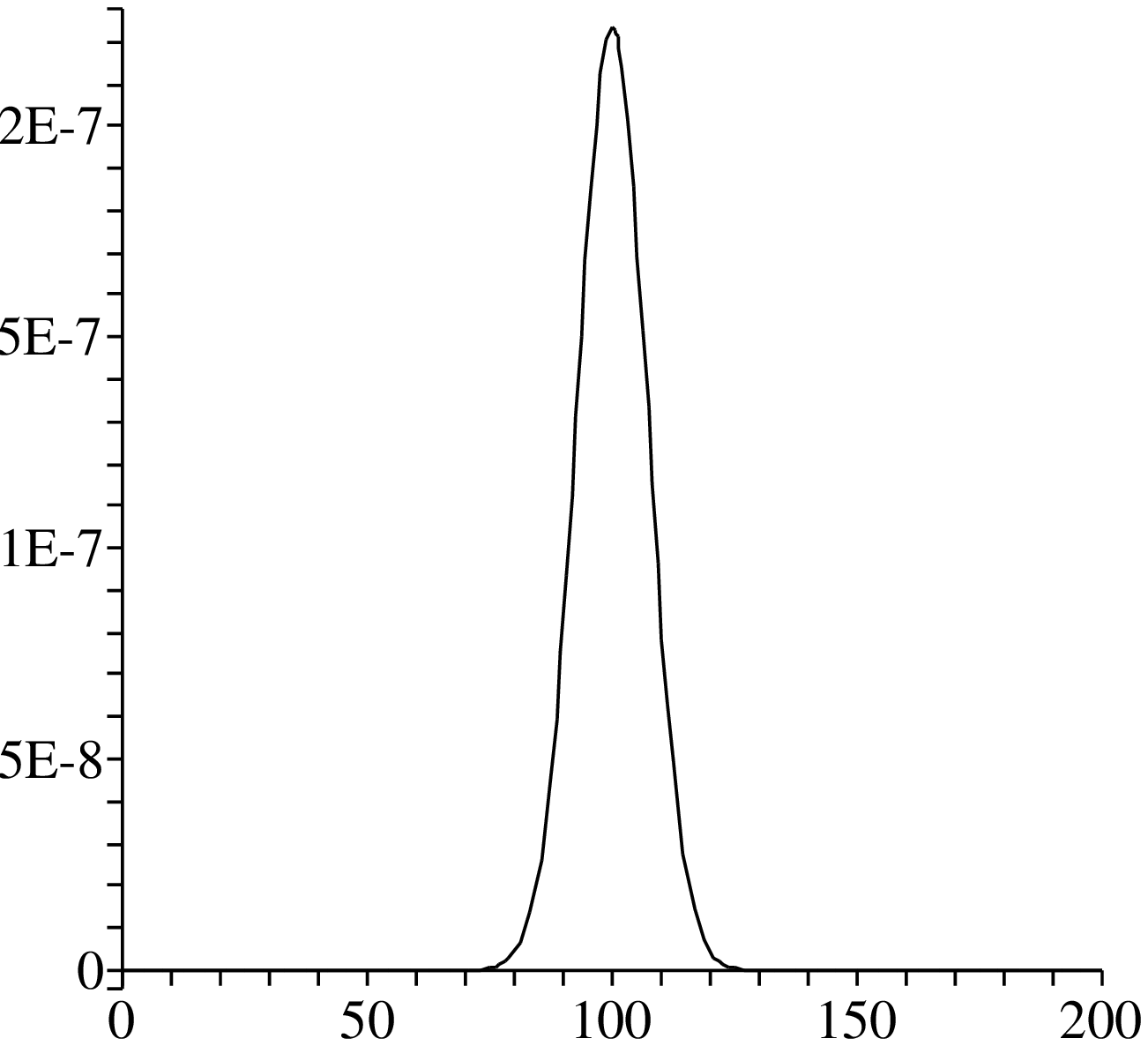}}
\put(-110,0){a) $\rho_o$ vs $r$}
\end{minipage}
\begin{minipage}{6cm}
\resizebox{6cm}{6cm}{\includegraphics{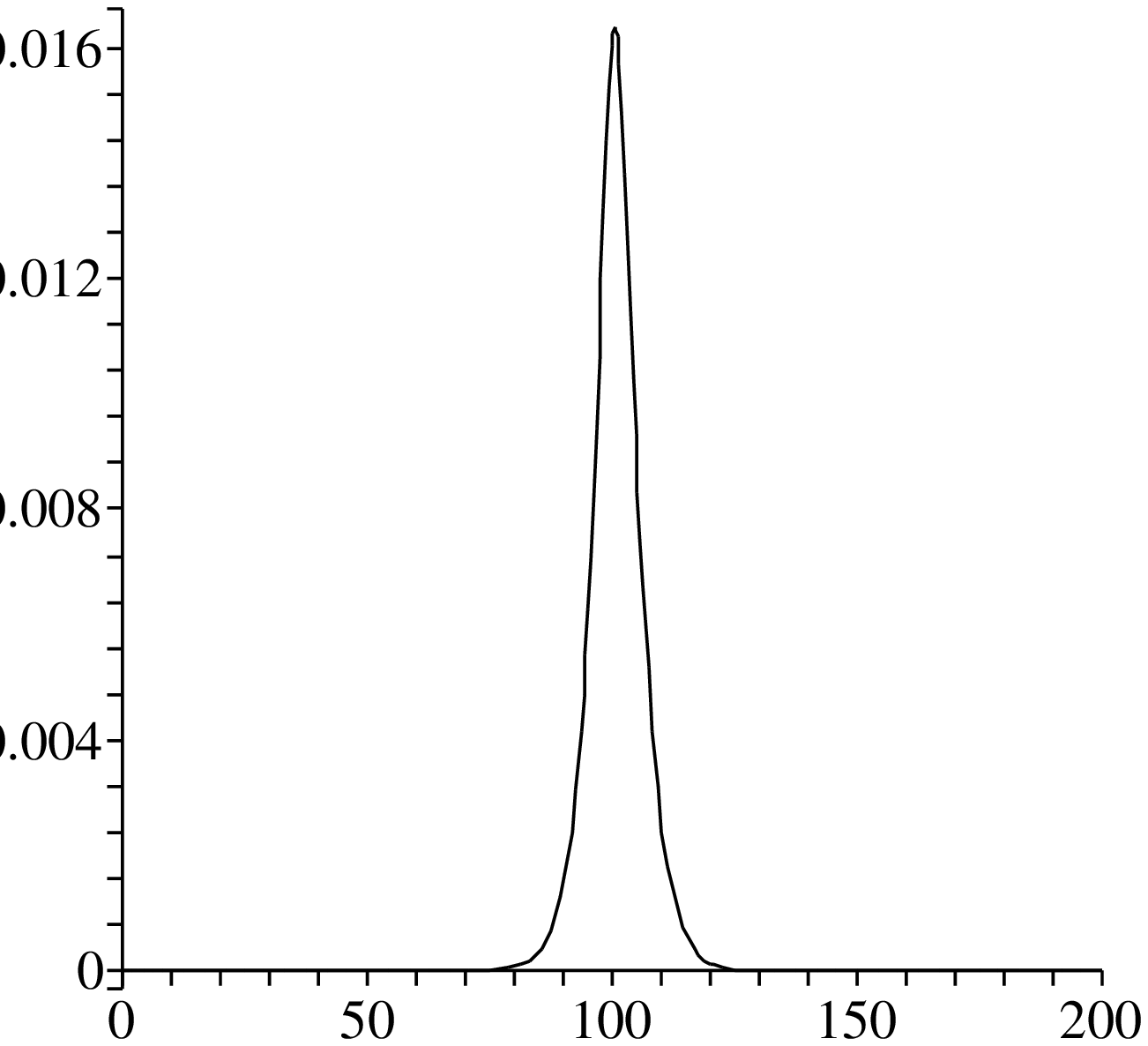}}
\put(-110,0){b) $C$ vs $r$}
\end{minipage}
\begin{minipage}{8cm}
\resizebox{8cm}{8cm}{\includegraphics{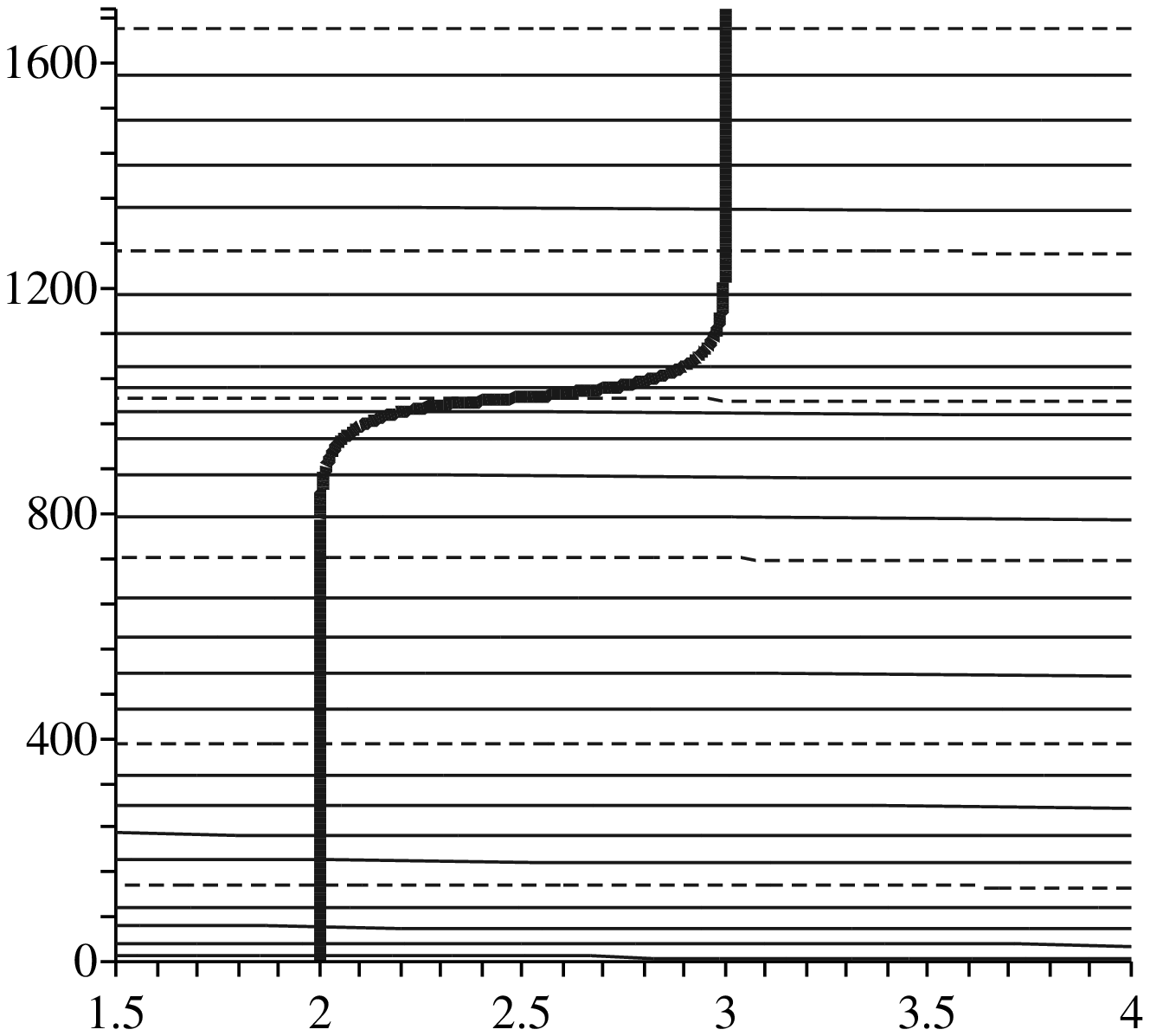}}
\put(-170,0){c) $\tau$ vs $R$ for MTT evolution}
\put(-25,35){ {\footnotesize{$r=25$}}}
\put(-25,63){ {\footnotesize{$r=50$}}}
\put(-25,98){ {\footnotesize{$r=75$}}}
\put(-25,128){ {\footnotesize{$r=100$}}}
\put(-25,156){ {\footnotesize{$r=125$}}}
\put(-25,199){ {\footnotesize{$r=150$}}}
\end{minipage}
\caption{Small dust shell}
\label{small}
\end{figure}

We begin with a dust shell of the form 
\be
\rho_o = \frac{m_o e^{-\left( \frac{r}{r_o} 
- \alpha \right)^2}}{2 \pi^{3/2} 
(1 + 2 \alpha^2) r_o^3} \, ,
\ee
where $m_o$ measures the total mass of the shell (ie.\ the asymptotic 
behaviour of the associated mass function), $r_o$ characterizes its 
``thickness", and $\alpha$ gives its initial position in terms of $r_o$. 
Again this distribution is Gaussian, but this time it is a shell with 
peak density at $\alpha r_o$. We wish to study the accretion of such a 
shell onto a pre-existing black hole and so excise a small region in the 
interior of the shell spacetime and replace it by a Schwarzschild 
geometry with mass parameter $M$, putting the junction at some 
$r =\hat{r} > 2M$. From the discussion in subsection \ref{TBsols}, for $r > \hat{r}$ the 
mass function must then be  $m(r) = M + 4 \pi \int_{\hat{r}}^r \rho_o(\bar{r}) \bar{r}^2 d\bar{r}$. 

An example of such a  spacetime is shown in Figure \ref{small}, where we 
have chosen $\hat{r}=2.5\,M$, $m_o = M/2$, $\alpha = 10$, and 
$r_o = 10\,M$.  Note that the 
parameters cannot be chosen arbitrarily in this case as shell crossings 
can easily develop. In fact, a small increase of the mass parameter so that $m_o \approx 0.7\,M$ will 
be sufficient to cause these. 
That said, for the parameters that we have chosen this does not occur, and
we see that the horizon is initially quiescent and begins to expand as the dust falls 
through it. The expansion is spacelike (from 
figure\ref{dust}b{\footnotesize)}, $C>0$), peaks as the largest density 
of matter crosses the horizon, and then tails off along with the 
infalling dust. Asymptotically the horizon becomes null again with 
areal radius $R = 3\,M$, so that the mass function tends to 
$m = \frac{3}{2}M$.\footnote{The asymptotic values will in fact be 
slightly smaller than that; due to the excision at $r=\hat{r}$, the 
parameter $m_o$ is not \emph{exactly} equal to the mass of the shell. 
However, in this example the difference is negligible.} 

\begin{figure}
\input{SpaceTimeAccretion.pstex_t}
\caption{Partial Penrose-Carter diagram showing the accretion of a small dust shell onto a pre-existing black hole. To get the full diagram this one should be : 1) continued on its left side with the Schwarzschild solution corresponding the the initial mass and 2) reflected through the instant of time
symmetry. }
\label{STA}
\end{figure}
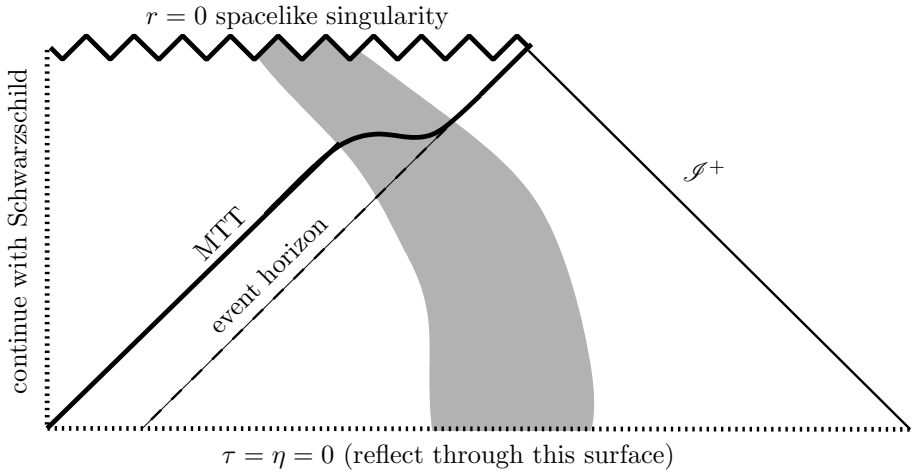

Figure \ref{STA} shows the corresponding spacetime diagram. From the spacetime surgery,  inside the dust shell the
spacetime is Schwarzschild with mass $M$. However, as the dust reaches and crosses the MTT, it begins to expand
in a spacelike manner and continues to do so as long as the dust continues to fall in. Ultimately as that density goes to zero the MTT asymptotes to the null event horizon. As in figure \ref{STC} the MTT clearly reacts to physical events while the event horizon, being teleologically defined, does not.  Note too that as in that earlier diagram, 
the dust  distribution is again, for simplicity,  shown with an edge.


\subsubsection{Large shell of approximately constant density falls into 
black hole}
\label{timelike}

\begin{figure}
\begin{minipage}{6cm}
\resizebox{6cm}{6cm}{\includegraphics{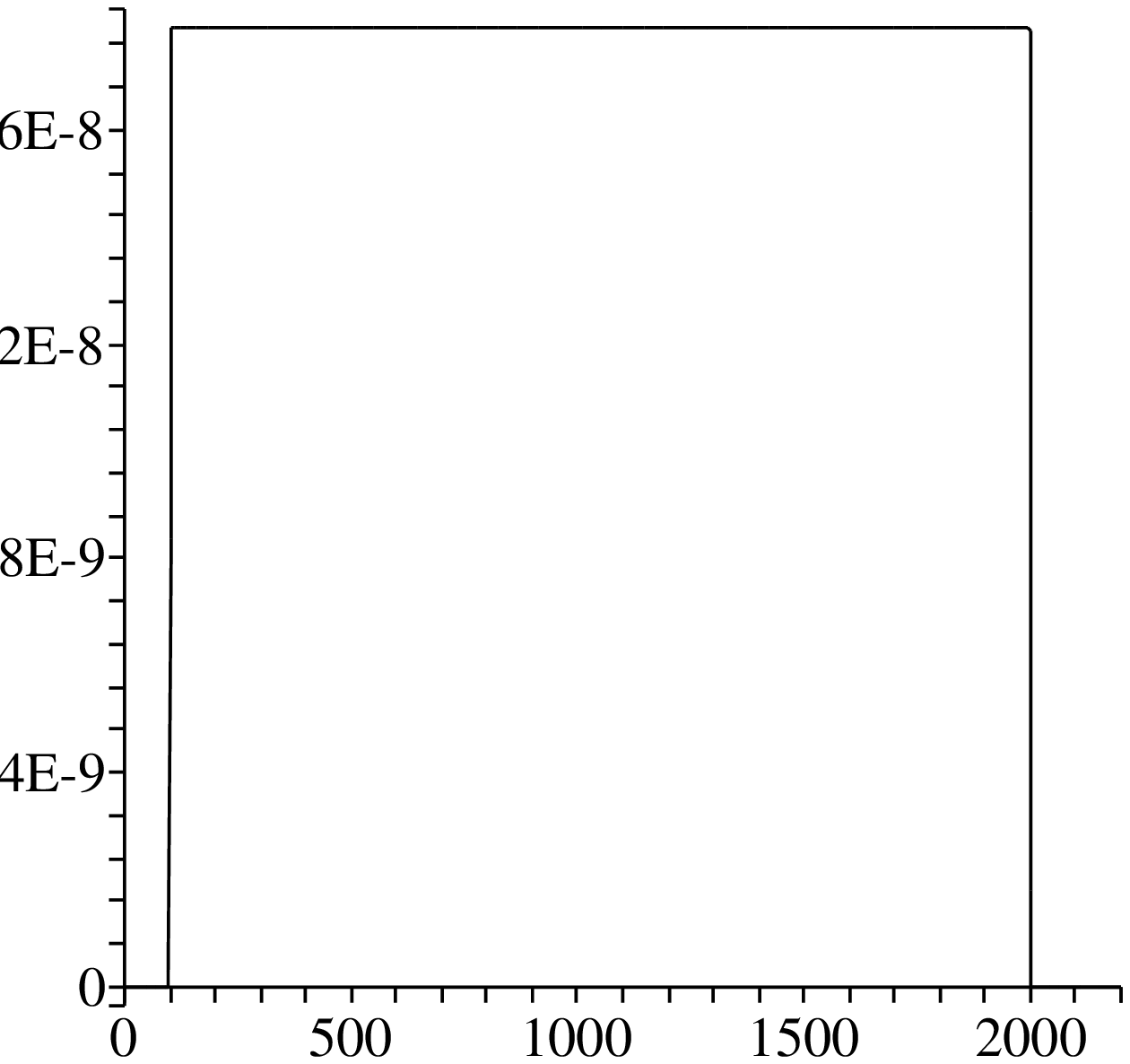}}
\put(-110,0){a) $\rho_o$ vs $r$}
\end{minipage}
\begin{minipage}{6cm}
\resizebox{6cm}{6cm}{\includegraphics{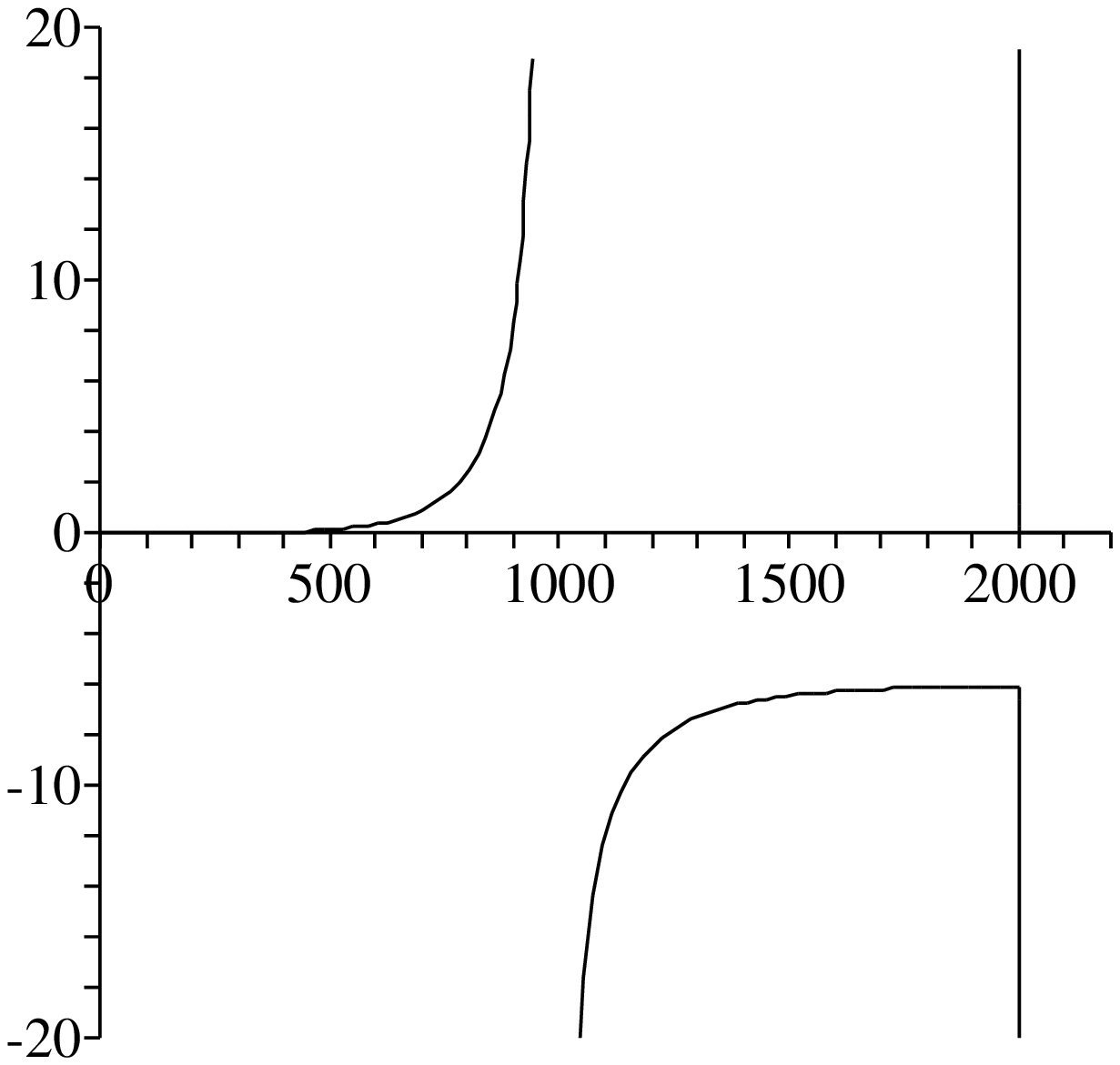}}
\put(-110,0){b) $C$ vs $r$}
\end{minipage}
\begin{minipage}{8cm}
\resizebox{8cm}{8cm}{\includegraphics{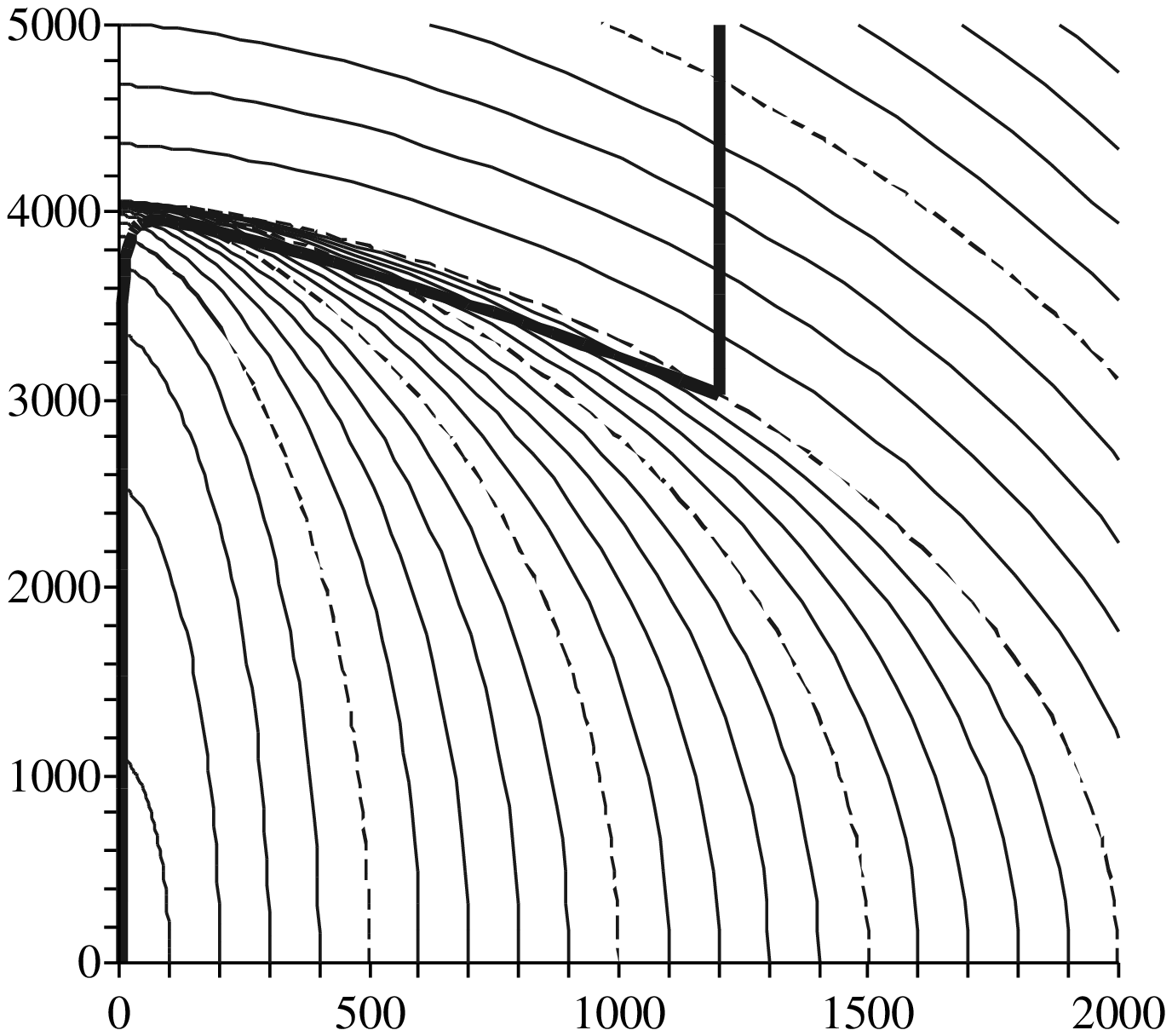}}
\put(-170,0){c) $\tau$ vs $R$ for MTT evolution}
\end{minipage}
\caption{Large shell of constant density}
\label{large}
\end{figure}

In the previous example, the expansion was everywhere spacelike.
It described a fairly dramatic situation, a significant expansion of a 
pre-existing black hole, however there was no sign of timelike membranes 
and the MTTs were isolated or dynamical horizons. We will now construct an 
example with timelike membranes. This will essentially be a smooth 
version of one of the more complicated Oppenheimer-Snyder examples 
presented in \cite{bendov04a}. Thus, we build a spacetime in which a 
large amount of (initially) constant density dust falls into a black hole. 
The initial density is given by
\be
\rho_o = 
\frac{3 m_o 
\left(\mbox{erf}(\frac{r - r_1}{M})  
- \mbox{erf}(\frac{r - r_2}{M})\right)}{4 \pi(r_2-r_1)(2 r_1^2+2r_1 r_2 
+ 2r_2^2 + 3 M^2)}  \,  , 
\ee
where $r_1$ and $r_2$ mark the (approximate) start and end of the shell, 
$m_o$ is the total mass of the shell as read from the mass function.
Excision is performed and the interior is replaced by a black hole
with mass $M$, which joins onto the dust exterior at some $r=\hat{r} > 2M$.
The full mass function for $r > \hat{r}$ will then be 
$m(r) = M + 4 \pi \int_{\hat{r}}^r \rho_o(\bar{r}) \bar{r}^2 d\bar{r}$.

For this example we choose $r_1 = 100\,M$, $r_2 = 2000\,M$, and 
$m_o = 600\,M$. Thus, the mass of the hole will increase dramatically 
during the evolution from $m=M$ to $m=601\,M$.  This evolution is shown in 
figure \ref{large}. Note that despite the appearance of the density 
function, it is actually smooth for $r > \hat{r}$ since the error 
functions themselves are smooth. 
Further, the density function will ``spread'' as it falls towards the 
hole and so the dust will not be of constant density as it crosses the 
horizon. 

That said, we see that the horizon begins to expand in the usual way 
as the initial matter falls into it. 
Then however, something different happens. Considering evolution with respect 
to $\tau$ we see that around $\tau = 3000\,M$, a new marginally trapped 
surface appears at $R=1202\,M$ and bifurcates into a dynamical 
horizon and a timelike membrane.
The dynamical horizon quickly asymptotes to an isolated horizon while 
the timelike membrane contracts and 
eventually annihilates with the dynamical horizon that grows out to 
meet it. 
 
Alternatively if we consider evolution with respect to the initial areal 
radius $r$, the MTT starts off almost-isolated, becomes dynamical as the 
first mass falls in, then goes null (with $-n^a$ as a tangent) and 
becomes a timelike membrane that continues to expand as it travels 
backwards in time. Finally, as the last dust falls through, it again 
goes null (with $-n^a$ as a tangent), dynamical, and then asymptotes 
back towards isolation. 

The spacetime diagram for this evolution would be very similar to that shown in figure \ref{STA}. The only 
difference would be that, during its active phase, the MTT would sometimes be timelike and so have a slope
greater than $45^\circ$.



\subsection{More complicated collapse}

The previous examples display the basic behaviours of marginally 
trapped tubes --- spacelike expansions, creation from singularities, and 
timelike contractions/backwards-in-time-expansions (depending how one 
views the evolution). In this section, to get a feel for the possible 
range of evolutions,  we will consider examples generated from more 
complicated initial conditions that combine several of these behaviours. 


\subsubsection{Spacelike expansion/multiple horizons}
\label{semh}

\begin{figure}
\begin{minipage}{6cm}
\resizebox{6cm}{6cm}{\includegraphics{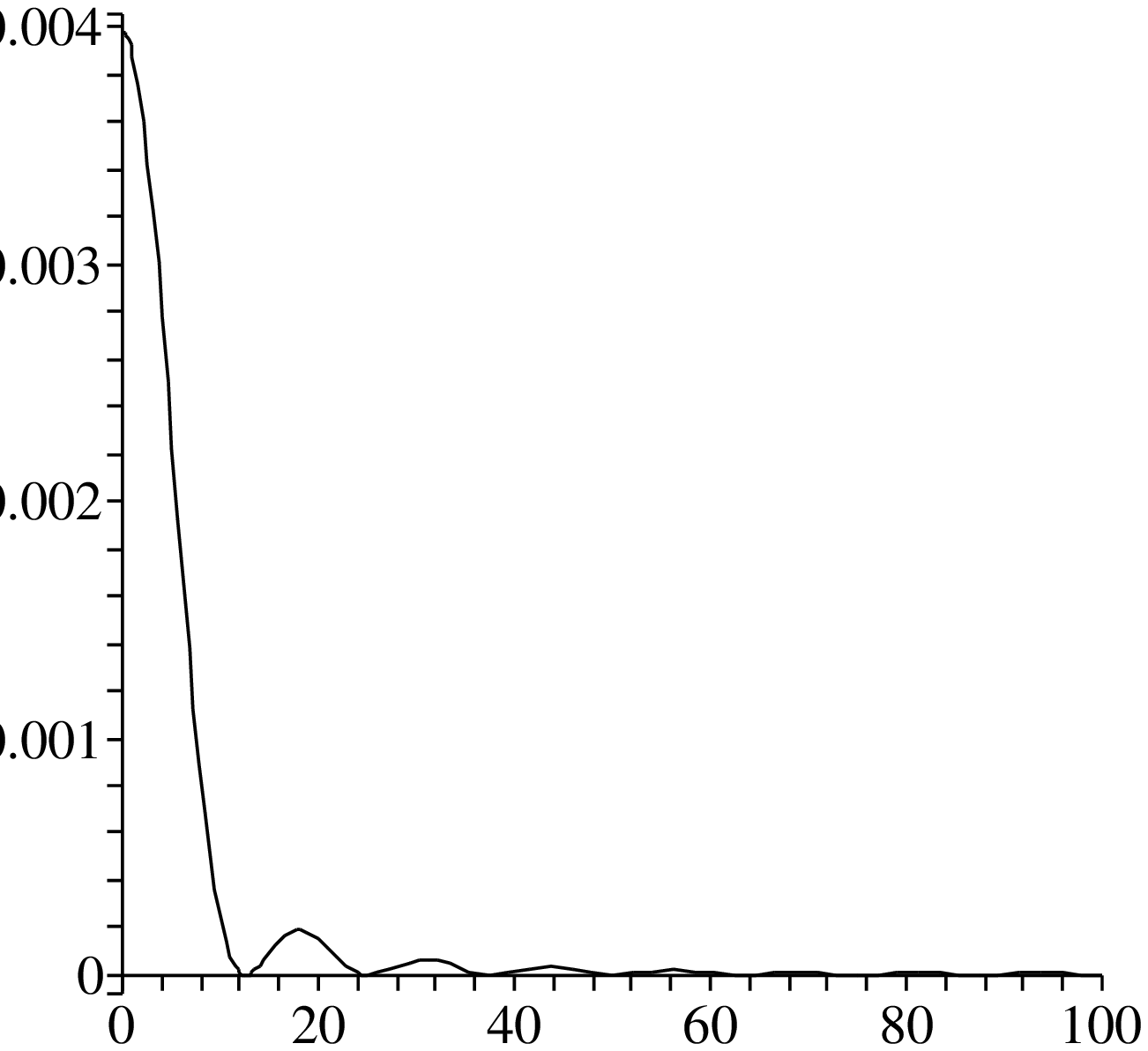}}
\put(-110,0){a) $\rho_o$ vs $r$}
\end{minipage}
\begin{minipage}{6cm}
\resizebox{6cm}{6cm}{\includegraphics{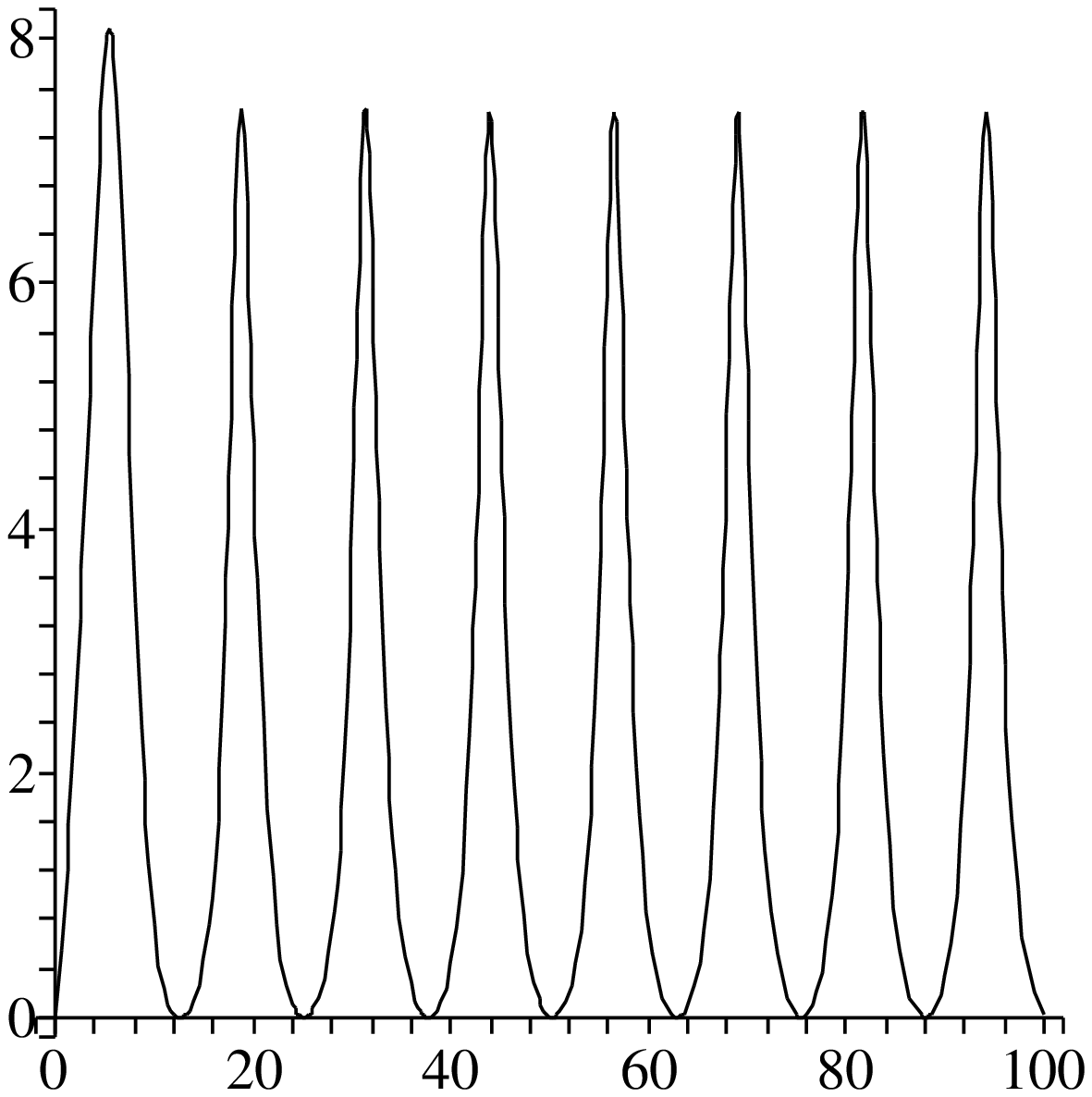}}
\put(-110,0){b) $C$ vs $r$}
\end{minipage}
\begin{minipage}{8cm}
\resizebox{8cm}{8cm}{\includegraphics{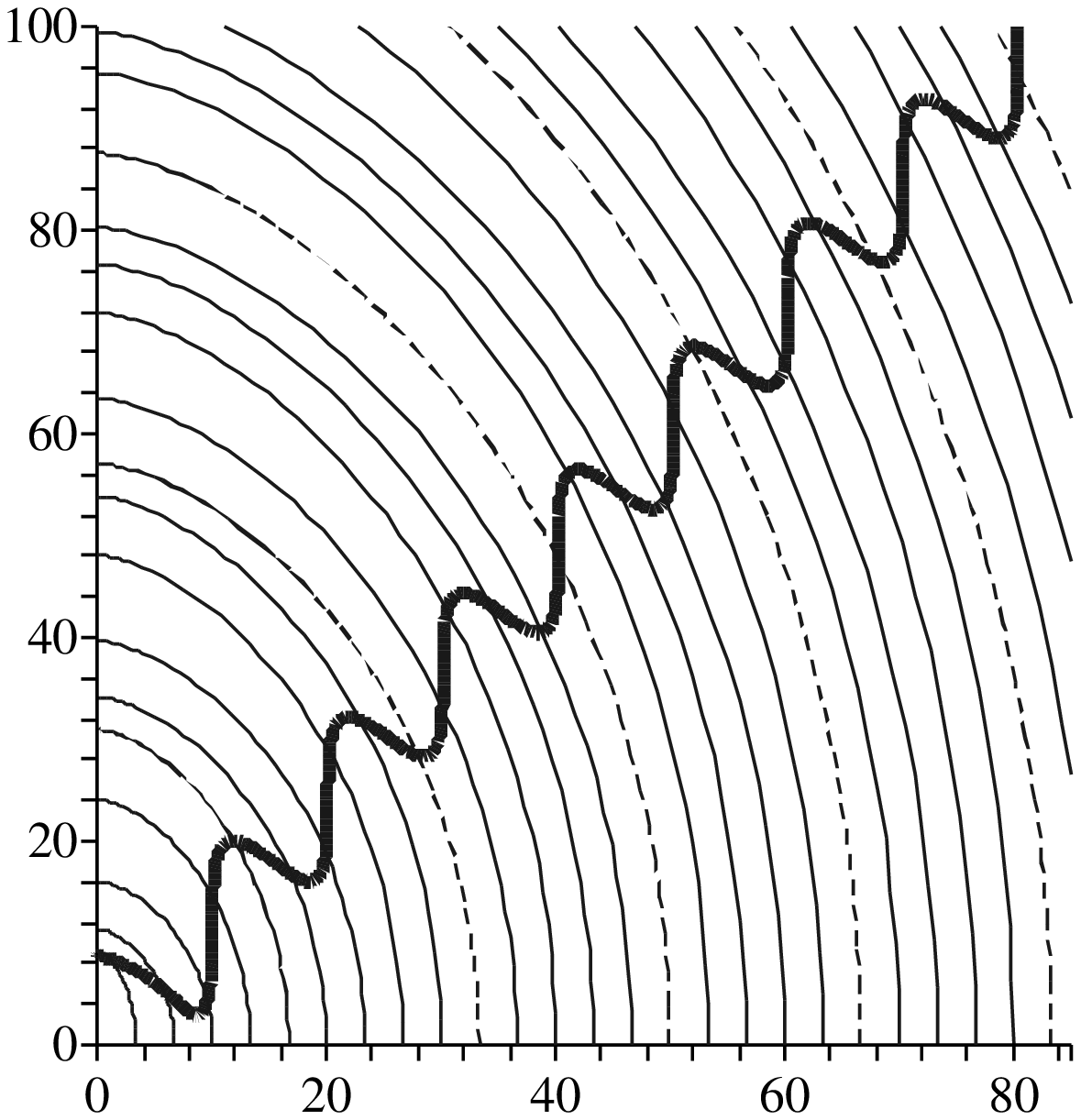}}
\put(-170,0){c) $\tau$ vs $R$ for MTT evolution}
\end{minipage}
\caption{Multiple dust shells fall into a black hole}
\label{sinshell}
\end{figure}

This example elaborates on a behaviour that we noted in section 
\ref{osexamples}. Namely it shows that the appearance of multiple 
horizons in a given leaf of a spacetime foliation does not always signal 
the existence of timelike membrane sections of the MTT. That is, a 
dynamical horizon as a spacelike surface can interweave with a foliation 
(of other spacelike surfaces) in highly non-trivial ways. 

We will consider matter distributions of the form
\be
\rho_o =  \frac{\alpha \mu}{2 \pi^2 r_o r^2} 
\sin^2 \left(\alpha\frac{r}{r_o}\right) \, ,
\ee
where $r_o$ is an arbitrary reference length scale. When this is 
integrated to a mass function, we see that for any positive integer 
$N$ there is an amount of mass $\mu$ between $r = N\pi r_o/\alpha$ and 
$r = (N+1)\pi r_o/\alpha$. Thus, physically this corresponds to a series 
of shells each with the same mass (though decreasing density). Again one 
has to choose parameters with care to avoid shell crossings and/or 
initial black holes. A particular choice that meets these criteria is
$\mu = (8\pi/5)\,r_o$ and $\alpha=1/4$. This is the distribution whose 
initial configuration and evolution is shown in figure \ref{sinshell}. 

Then, from Figure \ref{sinshell}c{\footnotesize)}, it is clear that 
$\tau = \mbox{constant}$ surfaces in this spacetime will contain either 
one or three marginally trapped surfaces (except at turning points 
of the MTT, where they contain two). Further, these will expand
and contract in apparently the same kinds of ways that we have seen 
previous example which include timelike membranes. However, an examination 
of Figure \ref{sinshell}b{\footnotesize)}, shows that despite this 
behaviour, the expansion parameter $C$ is always greater than zero and 
so the MTT is everywhere either spacelike or null/isolated. Then the 
apparent contractions/expansions arise simply because the MTT intersects 
the foliation in non-trivial ways. 

In Figure \ref{sinshell}c{\footnotesize)}, the horizon goes 
vertical/null at intervals of $R = 2 \mu \approx 10\,r_o$. Each of these 
corresponds to the density going to zero between each shell of mass 
$\mu = (8 \pi/5)\,r_o \approx 5\,r_o$ and so the horizon becoming 
instantaneously isolated. 
Finally, note that with a careful choice of the parameters, 
the newly created MTT can absorb shells of this type for arbitrarily long periods of time,  
always alternating between being dynamical and isolated. Thus, a 
horizon can absorb an arbitrarily large amount of mass without ever going 
timelike. It is the rate of absorption, not the total amount that is 
significant in this regard. 


\subsubsection{Multiple timelike membranes}
\label{mtlm}

\begin{figure}
\begin{minipage}{6cm}
\resizebox{6cm}{6cm}{\includegraphics{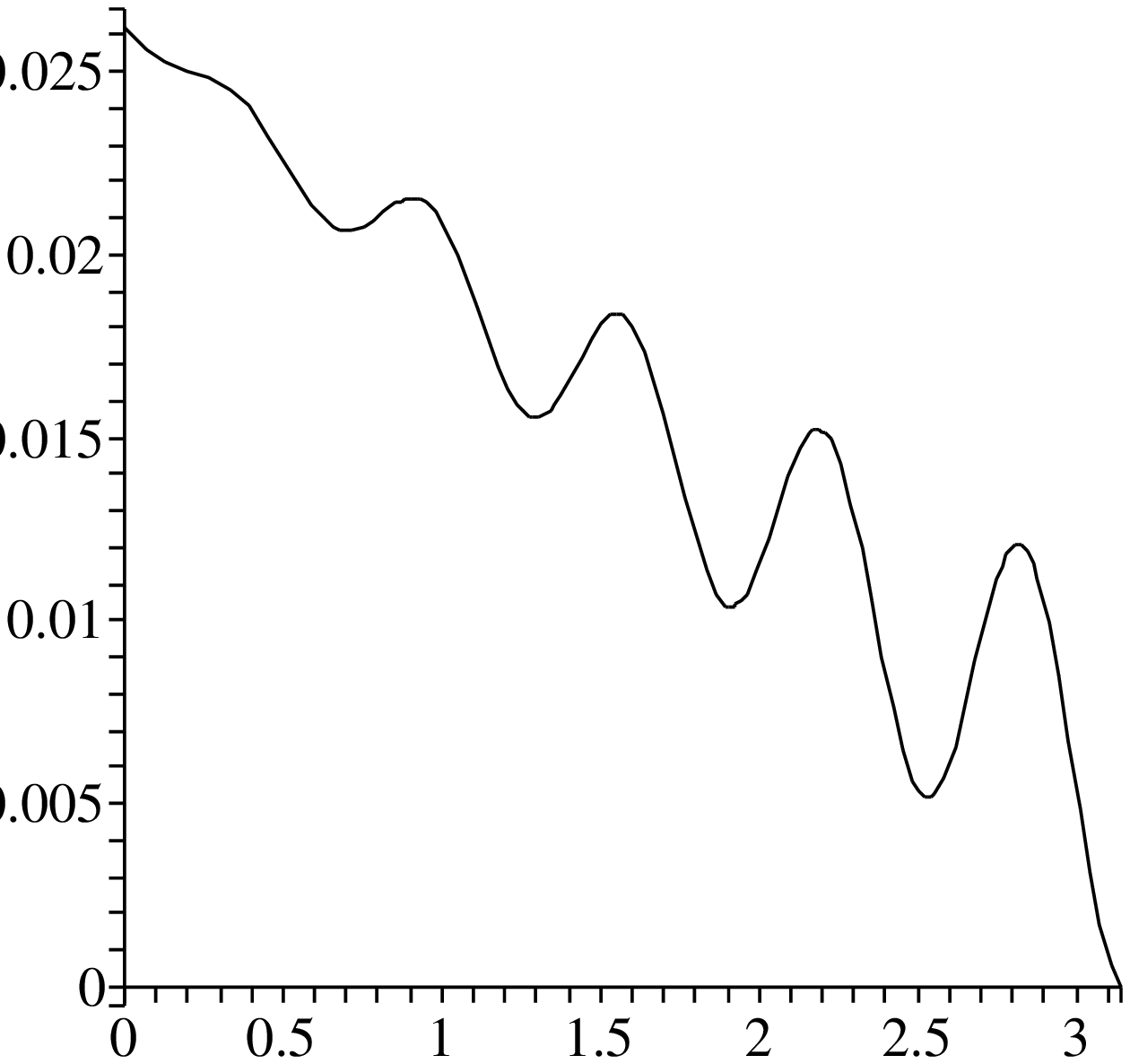}}
\put(-110,0){a) $\rho_o$ vs $r$}
\end{minipage}
\begin{minipage}{6cm}
\resizebox{6cm}{6cm}{\includegraphics{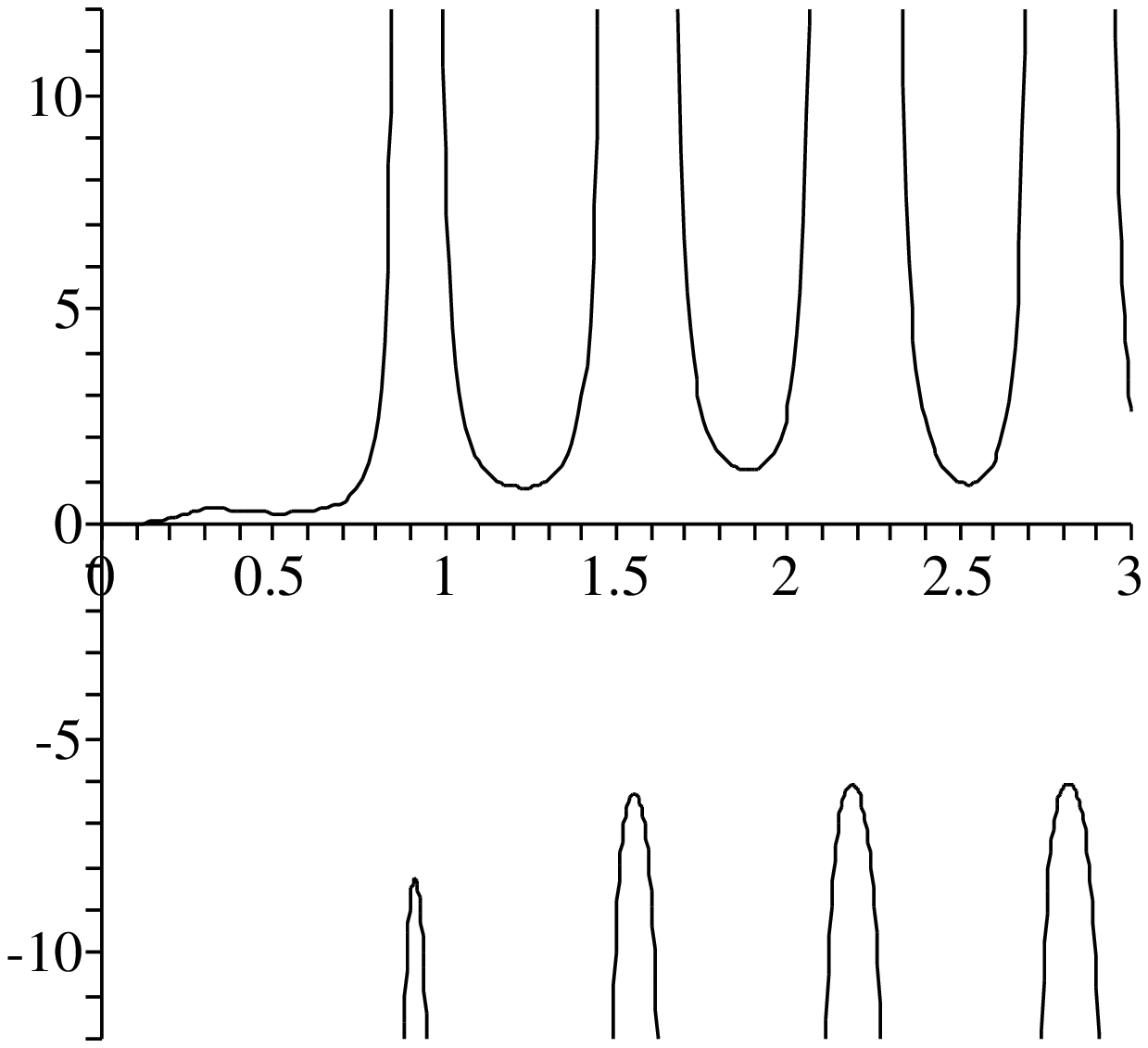}}
\put(-110,0){b) $C$ vs $r$}
\end{minipage}
\begin{minipage}{8cm}
\resizebox{8cm}{8cm}{\includegraphics{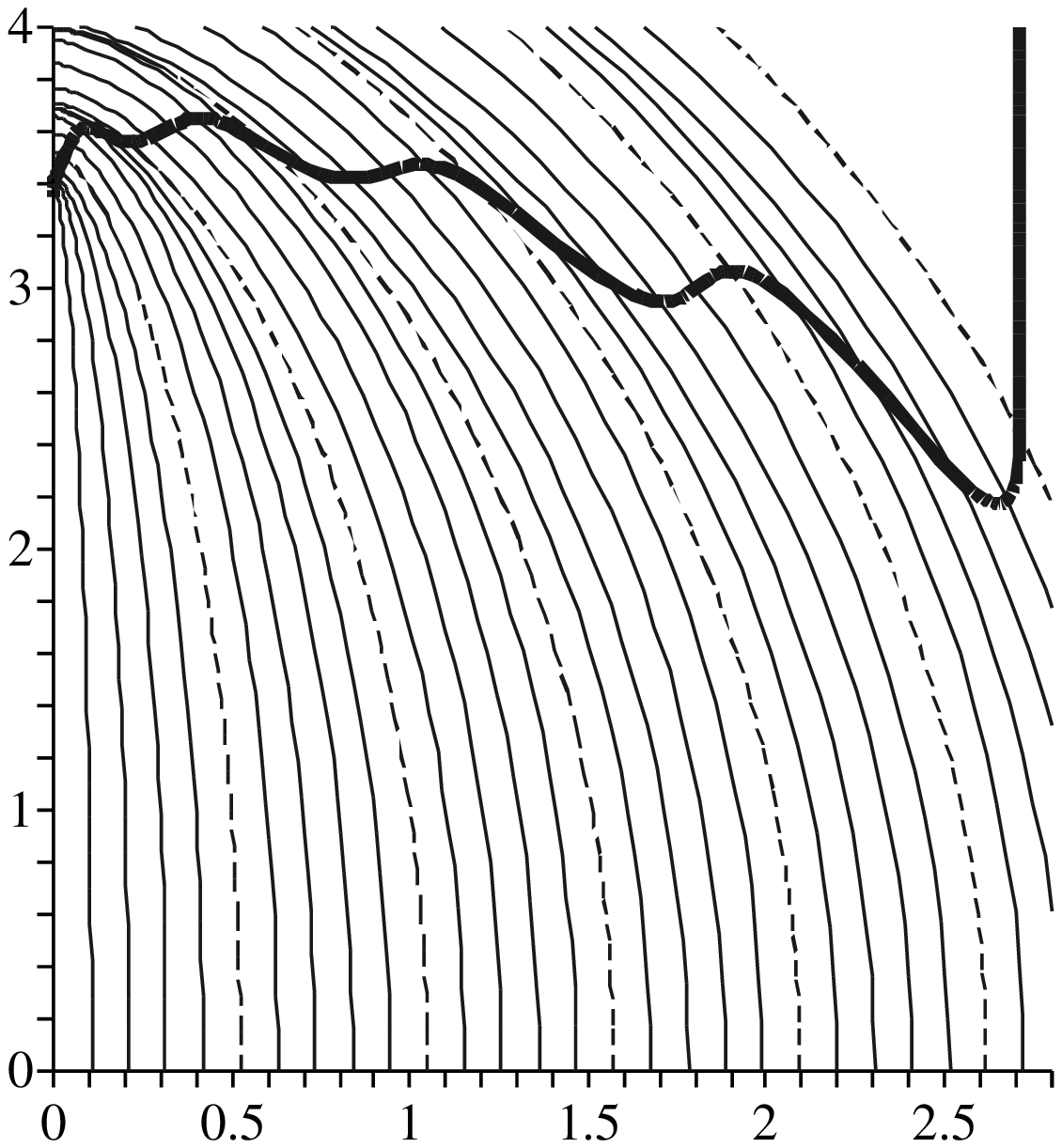}}
\put(-170,0){c) $\tau$ vs $R$ for MTT evolution}
\end{minipage}
\caption{A more complicated dust collapse.}
\label{R3}
\end{figure}

Our final example will demonstrate a spacetime that contains a MTT made 
up of multiple dynamical horizon and timelike membrane regions. The dust 
density will take the form:
\be
\rho_0 = \left\{ 
\begin{array}{ll}
\frac{\alpha}{r_o^2}\left[\pi 
-\frac{1}{5}\frac{r}{r_o}
\left(3+2 \cos^2 \left(5\frac{r}{r_o}\right)\right)\right] 
& 0 \leq r \leq \pi r_o \\ \\
0 & r >  \pi r_o
\end{array} \right.
\ee
where $\alpha$ is a dimensionless constant.
The exterior of $\pi r_o$ was excised to avoid negative density dust and
so violations of the energy conditions. Thus, outside of that radius, the geometry
will be Schwarzschild with mass $M = m(\pi r_o)$. 

Taking $\alpha = 1/120$ there are neither shell-crossings nor 
initial black holes in this spacetime. 
The evolution is then shown in the graphs of figure \ref{R3}. The initial 
density is irregular and gives rise to an MTT which, if we think of 
evolution as parameterized by $r$, starts out as a dynamical horizon and 
then alternates back and forth between being timelike and spacelike with 
(non-isolated) null cross-sections separating these regions. From the 
point of view of evolution with respect to $\tau$, spacelike slices have 
anywhere between one and five marginally trapped surfaces in this example. 
However, all of these intricacies are contained within the 
outermost isolated horizon (and here it really is isolated since we cut 
the density distribution at $r=\pi r_o)$ and would not be visible to 
outside observers. 


\section{Scalar fields}
\label{scalarfields}

In the previous section we have seen that the potential MTT behaviours 
suggested by (\ref{perfectfluid}) and (\ref{heuristic}) are all achieved 
by the dust spacetimes. Thus, depending on the initial configuration, the 
MTT can be any of spacelike, null, or timelike. In this section we will 
attempt the same demonstration for the scalar fields. Thus, we will 
consider initial configurations of scalar fields, evolve them in time, 
and examine the behaviour of any MTT that forms with the help of the
expansion parameter $C$. 

Given that scalar fields are significantly more complicated than dust
we will necessarily take a numerical rather than an analytic approach.\footnote{
There are a few exact scalar field
solutions in the literature, see for example \cite{viqar} and the references listed therein. 
However, they are much more restricted than the Tolman-Bondi solutions 
(for instance, we are not aware of any that are asymptotically flat) and cannot produce
the same variety of examples.} Section~\ref{sec:num_app} will introduce 
the numerical model and the subsequent sections will present the results 
for two different scalar fields configurations. These will be analogous to
configurations considered in the last section. The first, in 
section~\ref{sec:step}, examines the evolution of a initially smooth 
step-like configuration that collapses to form a black hole. The second, 
in section~\ref{sec:shell}, studies a shell of scalar field which falls 
into an existing black hole. 
Due to numerical complexities, these results will be less complete than 
those considered in the last section, but will still demonstrate some 
interesting and complementary  behaviours. 


\subsection{Numerical approach}
\label{sec:num_app}

We will be interested in the scalar fields briefly considered in section \ref{scalar}. 
The equations governing spacetimes containing these fields are generated by the Lagrangian:
\begin{equation}
{\cal L} = \sqrt{-g} \left[  \frac{1}{16\pi} R 
- \frac{1}{2}  \nabla_\alpha \phi \nabla^\alpha \phi 
- V(\phi) \right]  \; .
\label{eq:lag} 
\end{equation}
This class of models includes the massive Klein-Gordon field with 
$V=m_0 \phi^2/2$ and in the following we will restrict ourselves to this 
case. 

To solve the resulting coupled Einstein-Klein-Gordon equations we work 
with the standard 3+1 approach based on the ADM 
equations~\cite{ADM} and in particular adopt the techniques 
introduced in~\cite{Regular} to perform the evolutions. 
Restricting to spherical symmetry, we can then study the general 
evolutions by considering the dynamics of spacetimes with metrics of 
the form 
\begin{equation}
ds^2 = - \alpha^2(r,t) dt^2 + A(t,r) dr^2 + r^2 B(r,t) \left(d\theta^2 
+ \sin^2 \theta d\varphi^2 \right) \; . 
\label{eq:metric}
\end{equation}
The spherical symmetry implies that all dynamical functions depend
only on $r$ and $t$. $\alpha$ is the usual lapse function and for 
simplicity  we impose a vanishing-shift gauge condition; equivalently 
the ``time-evolution" vector, $\partial/\partial t$ is everywhere 
orthogonal to the $t = \mbox{constant}$ ``instantaneous" three-surfaces. 

Focusing on this natural foliation with respect to $t$, we note that 
these hypersurfaces have intrinsic three-metric and extrinsic curvature :
\begin{eqnarray}
h_{ab} &=& A [dr]_a [dr]_b + B r^2 \Omega_{ab} \; \; \mbox{and}\\
K_{ab} & = & \frac{1}{2 \alpha} \left( \frac{\partial A}
{\partial t} \right) [dr]_a [dr]_b 
+  \frac{1}{2 \alpha} \left( \frac{\partial B}{\partial t} \right) 
\Omega_{ab} \, ,
\end{eqnarray}
respectively, where $\Omega_{ab}$ was defined following equation 
(\ref{TBK}). These, together with the value of the scalar field, are the 
variables whose evolution we will study.  

Then, in the usual way we can rewrite the Einstein-Klein-Gordon equations 
in terms of the Hamiltonian and momentum constraints which restrict 
initial values of these quantities, in addition to evolution equations 
(which preserve the constraints). 
The actual implementation consists of picking an initial configuration 
for the scalar field, solving the constraints for $h_{ab}$ and 
$K_{ab}$ and finally using the evolution equation to obtain the dynamics 
of the corresponding spacetime. For brevity we will not include all of 
the details here; the interested reader is directed to 
\cite{Regular} for a more detailed discussion.

Once we have a spacetime evolution, the next step is to search for an 
MTT. In this case it is natural to search for apparent 
horizons/marginally trapped surfaces within the $t = \mbox{constant}$ 
slices and this is what we will do. Since we are in spherical symmetry 
we need only consider the null expansions of the $r = \mbox{constant}$ 
two-surfaces. 
Then, with 
\be
u_a = - \alpha [dt]_a  \; \; \mbox{ and } \; \;  s_a 
= \sqrt{A} [dr]_a \, , 
\ee
as the forward-in-time and towards-infinity pointing unit normals to 
the two-surfaces, we define
\be
\ell_a = u_a + s_a \; \; \mbox{ and } \; \;  n_a = \frac{1}{2}(u_a - s_a)  
\ee
as the outward and inward forward-in-time pointing null normals to those 
same surfaces. As in the dust examples,  the exact ``normalization'' is 
irrelevant as we are only interested in the signs of the ensuing 
quantities, not their magnitudes. Thus, we can make this specific, 
convenient, choice. Then, it is straightforward to see that:
\begin{eqnarray}
\thl &=& {\cal D}_a s^a  + K - K_{ab} s^a s^b  =   
 \frac{\partial_t B}{\alpha B }+
\frac{1}{A^{1/2}} \left( \frac{2}{r} + \frac{\partial_r B}{B} \right) \\ 
\thn &=& - \frac{1}{2}({\cal D}_a s^a -K + K_{ab} s^a s^b)
= \frac{\partial_t B}{2 \alpha B } - 
 \frac{1}{2A^{1/2}} \left( \frac{2}{r} + \frac{\partial_r B}{B} \right) 
\, , 
\label{eq:expansion}
\end{eqnarray}
where ${\cal D}_a$ is the covariant derivative compatible with 
$h_{ab}$ and $K = h^{ab} K_{ab}$ (as a quick check that these 
expressions are reasonable, note that for Minkowski space 
where $A=B=1$, $\thl = 2/r > 0$ and $\thn = - 1/r < 0$).
In the code we evaluate these expressions over the whole numerical grid 
and look for places where $\thl$ changes signs and $\thn < 0$. The 
corresponding two-spheres are then identified as the marginally trapped 
two-spheres which foliate the MTTs, and their area is 
\begin{equation}
A_{AH} = 4 \pi r_{AH}^2 B_{AH} \;
\end{equation}
where $B_{AH}$ is the metric function $B$ evaluated at \mbox{$r=r_{AH}$}, 
the coordinate radius of these two-spheres. 

Finally, the expansion parameter $C$ for this surface can be calculated 
using (\ref{Cscalar}).


\subsection{Scalar field ``step" collapses to form a black hole}
\label{sec:step}

\begin{figure}
\begin{minipage}{7.5cm}
\resizebox{7.5cm}{!}{\rotatebox{270}{\includegraphics{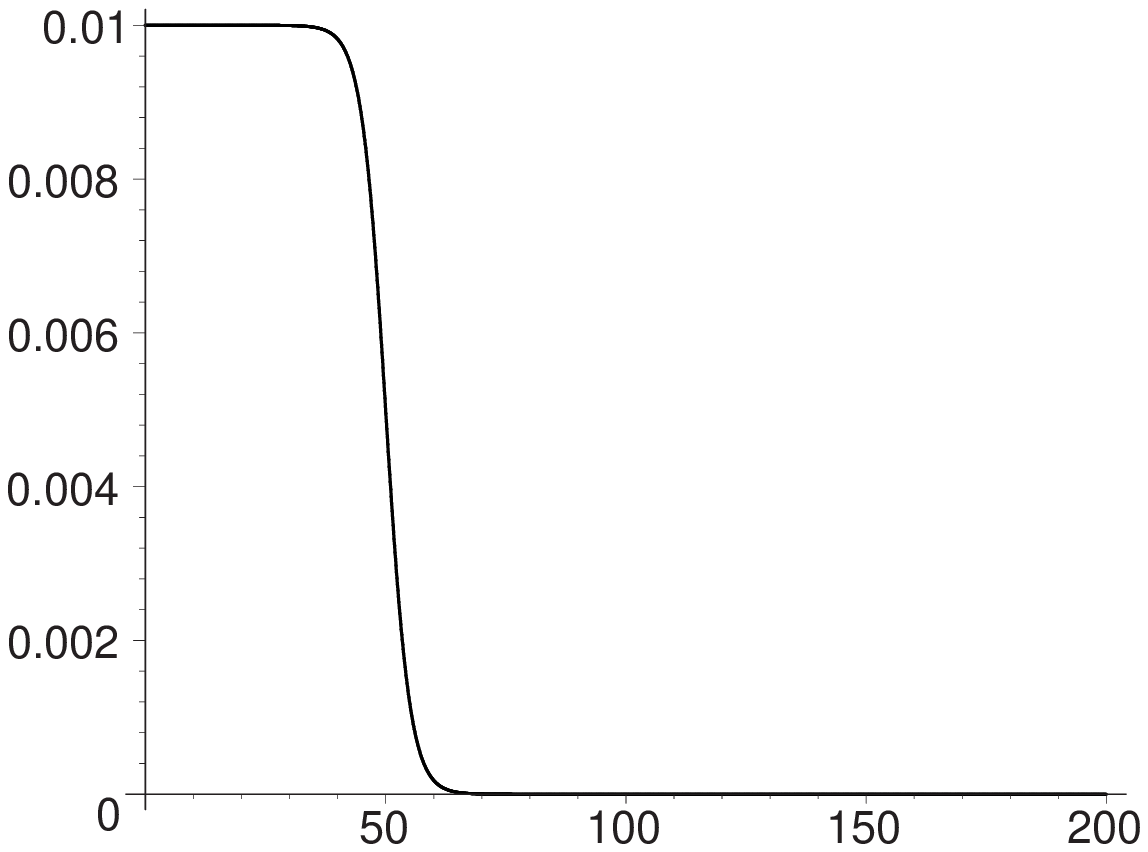}}}
\put(-140,-175){a) Initial $\phi$ vs $r$}
\end{minipage}
\begin{minipage}{7.5cm}
\resizebox{7.5cm}{!}{\rotatebox{270}{\includegraphics{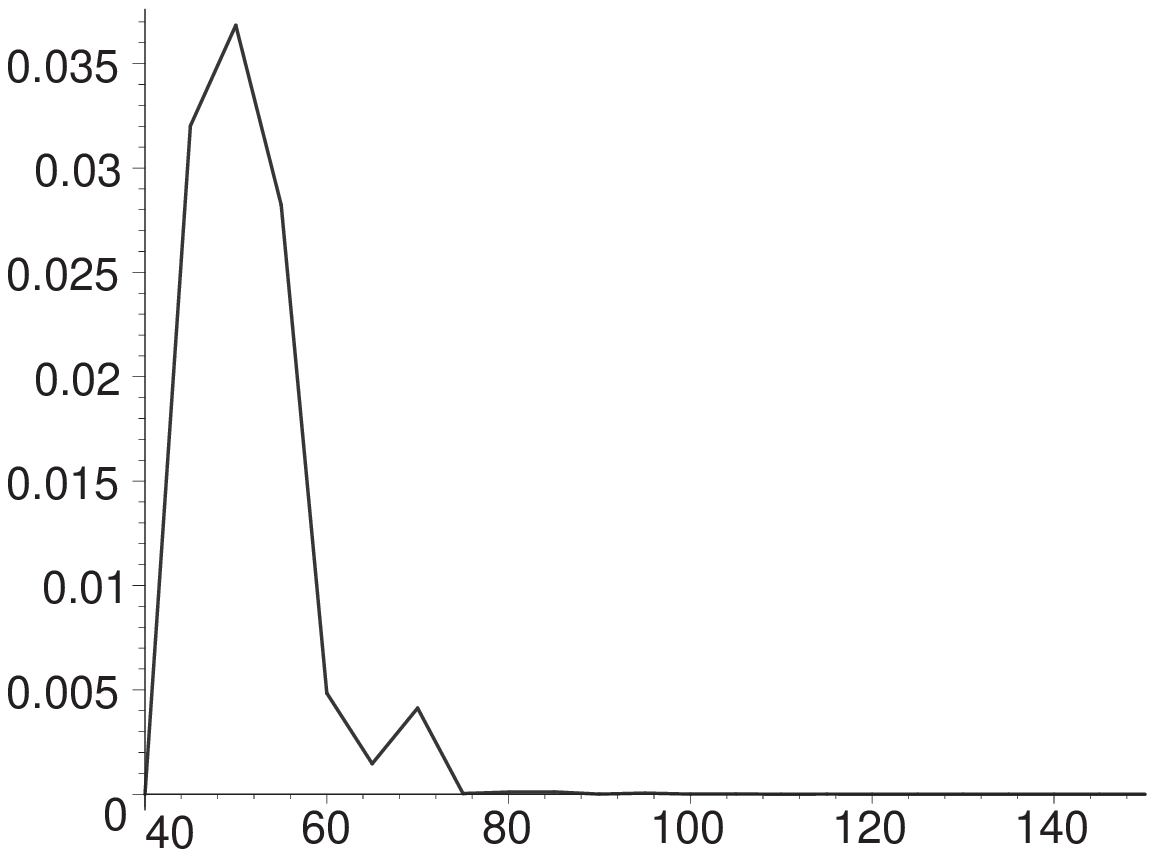}}}
\put(-140,-175){b) $C$ vs $t$}
\end{minipage}
\begin{minipage}{9cm}
\resizebox{9cm}{!}{\rotatebox{270}{\includegraphics{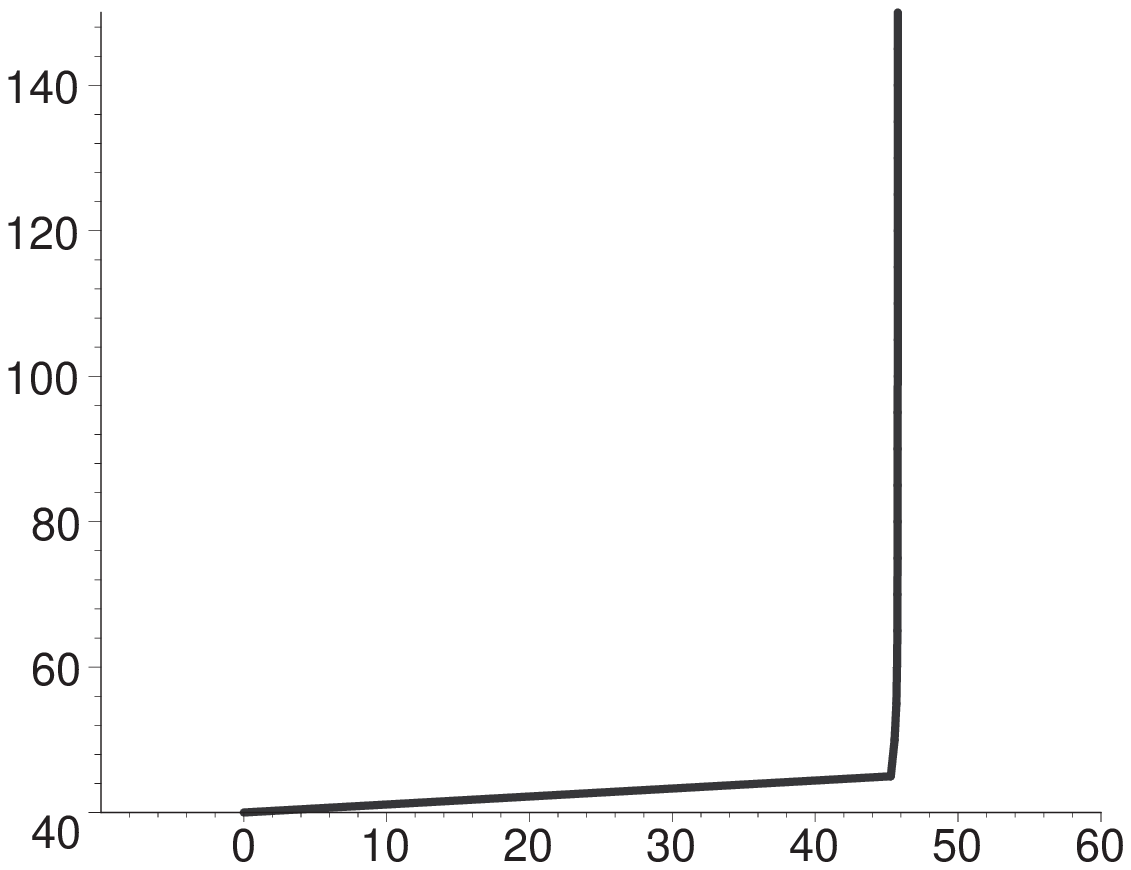}}}
\put(-180,-210){c) $t$ vs $R$ for MTT evolution}
\end{minipage}
\caption{Scalar field ``step''}
\label{phiStep}
\end{figure}

Our first example will be analogous to the dust ball collapses considered 
in section \ref{dustcollapse_sec}. As for those dust examples, our initial 
slice will be a moment of time symmetry ($K_{ab} = 0$) and further we will 
choose our coordinates so that  the coordinate radius $r$ 
will also be the areal radius $R$ on that slice (thus $B=1$ initially). Then, we specify 
a step-like configuration of the scalar field as shown in 
figure \ref{phiStep}a{\footnotesize)}. 

Solving the constraint equations to find the initial form of $A$, we can 
then integrate the data as discussed above to find how the geometry and 
fields evolve in time. The results are shown in figures \ref{phiStep}b{\footnotesize)} 
and \ref{phiStep}c{\footnotesize)}. The first thing to note is that throughout this
collapse $C > 0$ and so the MTT is a dynamical horizon. Secondly, as in 
the earlier examples it asymptotes to null as the spacetime settles down 
to become Schwarzschild. 
Note from figure \ref{phiStep}c) that initially the hypersurfaces
do not contain an apparent horizon. However, during the collapse an 
apparent horizon appears and then keeps
growing until it reaches a constant size.


\subsection{Scalar field ``shell"  accretes onto existing black hole}
\label{sec:shell}

Our second scalar field example is analogous to the dust shell examples 
of section \ref{accrete}. Thus we will consider an initial ``shell" of 
scalar field that accretes onto an existing black hole of mass $M$. 
After surgically inserting the black hole we again start out on a slice of time symmetry, 
though in this case for technical reasons do not start with $r$ as the 
areal radius. 
Instead, on the initial slice $\tilde{B} =1$, where $B = \left( 1 
+ \frac{M}{2r}\right)^4 \tilde{B}$. 

\begin{figure}
\begin{minipage}{7.5cm}
\resizebox{7.5cm}{!}{\rotatebox{270}{\includegraphics{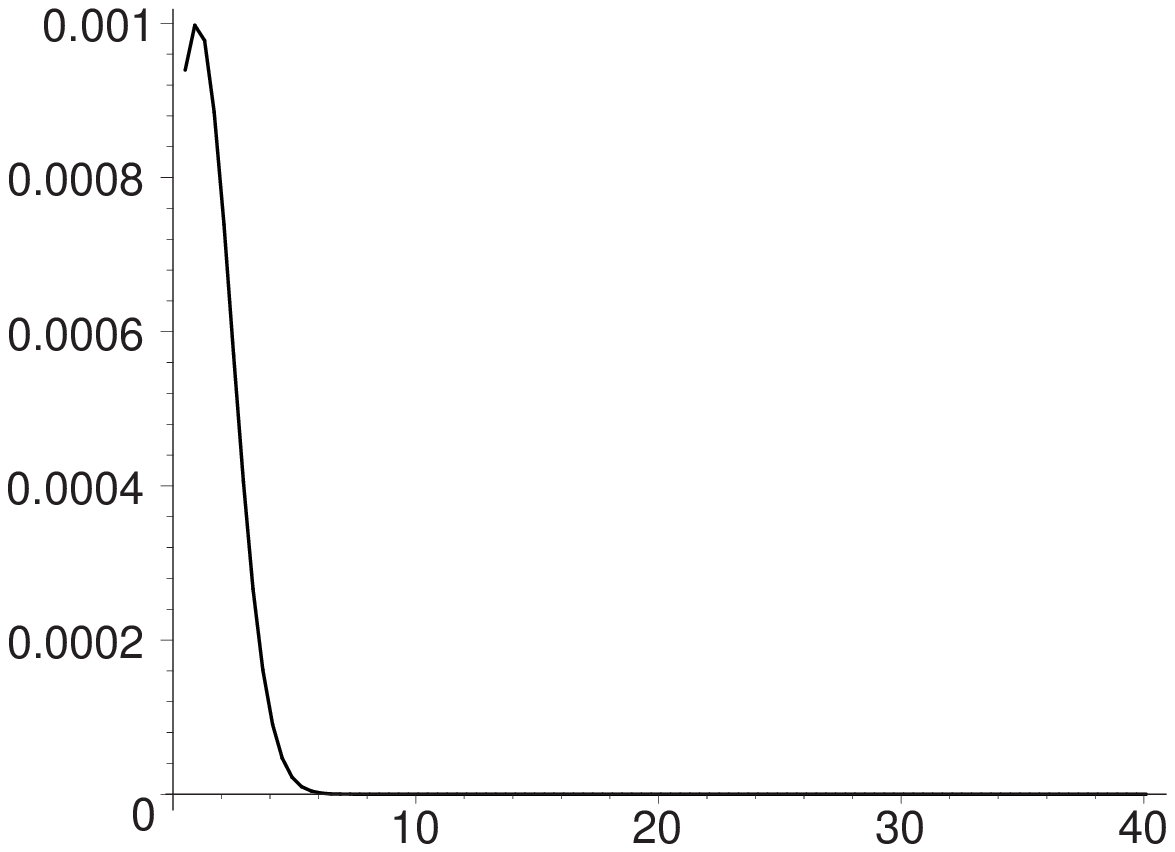}}}
\put(-140,-175){a) Initial $\phi$ vs $r$}
\end{minipage}
\begin{minipage}{7.5cm}
\resizebox{7.5cm}{!}{\rotatebox{270}{\includegraphics{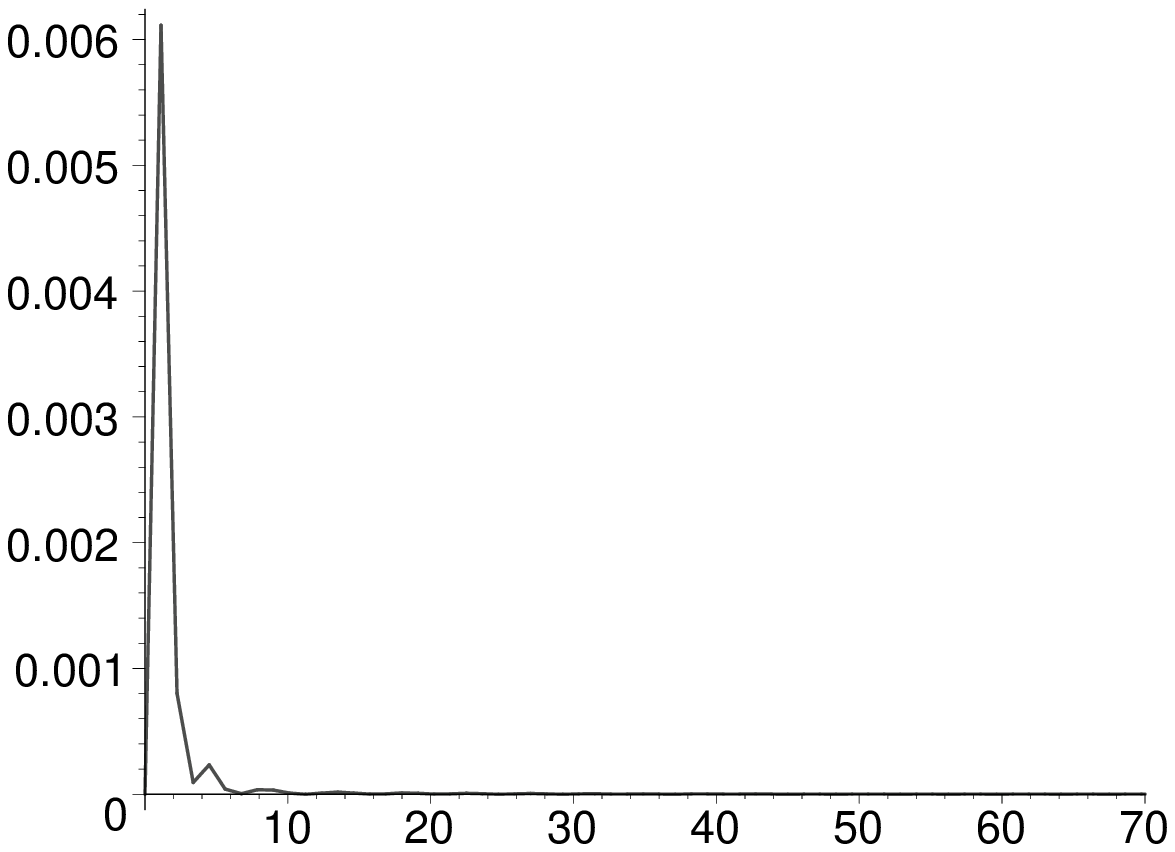}}}
\put(-140,-175){b) $C$ vs $t$}
\end{minipage}
\begin{minipage}{9cm}
\resizebox{9cm}{!}{\rotatebox{270}{\includegraphics{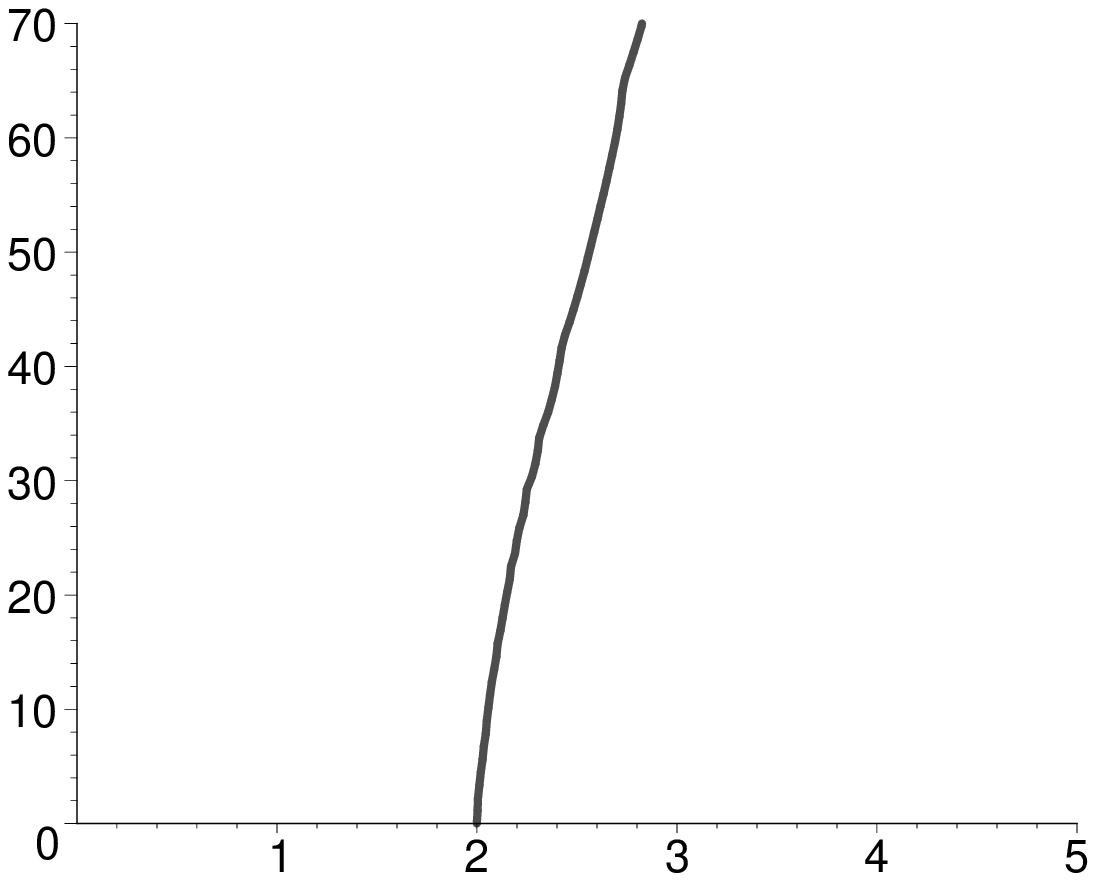}}}
\put(-180,-210){c) $t$ vs $R$ for MTT evolution}
\end{minipage}
\caption{Scalar field accretion}
\label{phiShell}
\end{figure}

Then, as our initial scalar field conditions we consider a $\phi(0,r)$ 
of the form shown in figure \ref{phiShell}a{\footnotesize)}. The corresponding initial 
intrinsic geometry (essentially only $A$ remains unknown) is 
then found by solving the constraint equations. 
The results for the corresponding evolution are then shown in figures 
\ref{phiShell}b{\footnotesize)} and \ref{phiShell}c{\footnotesize)}. On those graphs we again see that 
$C$ remains everywhere non-negative and asymptotes to zero at late times. 
Note too the apparent horizon on the initial hypersurface which grows as the scalar field falls into the hole. 


\subsection{Outlook for scalar fields}

Neither of the preceding examples included timelike membrane regions 
of the MTT. 
We believe that this is a reflection of the examples that our 
code has been able to integrate rather than a fundamental result. 
From (\ref{Cscalar}) it is clear that a timelike membrane will only appear in 
situations where there is a sufficiently large concentration of the 
scalar field relative to the (inverse) area of the horizon. Our code 
had difficulty evolving such examples and so the lack of timelike 
membranes is not especially surprising. Physically, one would also 
expect it to be more difficult to obtain such examples for a massive 
scalar field which, in contrast to pressureless dust, resists 
compression. We expect that future investigations will find the 
appropriate combination of initial conditions needed to generate 
examples of timelike membranes in scalar field spacetimes. 
For now though, we simply note that our examples show timelike evolutions are not the rule 
for scalar field spacetimes. 


\section{Conclusions and Speculations}
\label{conclusions}

In this paper we have seen that dynamical horizons and timelike membranes 
characterize two possible modes of black hole expansion. In the first  
a black hole/MTT smoothly expands as matter falls into it.
By contrast, the second occurs when matter densities are high enough to 
force the formation of a new horizon of 
non-zero area that encloses any already existing MTTs. 

Further insight into these two possibilities can be gained by considering 
the magnitudes of the quantities involved. As seen in the discussion 
of \ref{mtt}, assuming that the null energy condition holds, the signature of 
an MTT is determined by the relative magnitudes of $1/A$ and 
$2 T_{ab} \ell^a n^b$, or, specializing to (pressureless) dust, $1/A$ 
and $\rho$. Converting into physical units, it is straightforward to 
see that for a Schwarzschild black hole of mass $M = \mu M_\odot$, the 
inverse horizon area corresponds to a density
\be
\rho_A = \frac{c^6}{16 \pi G^3 M_\odot^2} \frac{1}{\mu^2} \, , 
\ee
where $M_\odot =  2 \times 10^{33}$ g is the approximate mass of the 
Sun and $c$ and $G$ are respectively the speed of light and the 
gravitational constant. Then, for a solar mass black hole, 
$\rho_A \approx 10^{16}\,\, \mbox{g}/ \mbox{cm}^3$ which is about an 
order of magnitude higher than the density of a neutron star. By contrast, 
for a supermassive black hole of mass $10^8\, M_\odot$, one has 
$\rho_A \approx 1\,\, \mbox{g}/ \mbox{cm}^3$, the density of water. 

Though strictly speaking these results only apply in situations of 
spherical symmetry, it seems fairly safe to use them to draw some more 
general conclusions. Specifically, it is likely that both modes of 
expansion are not only mathematically possible as we have seen in this 
paper, but also both occur in physical situations. For small black holes, 
even if the lack of spherical symmetry changes these estimates by several 
orders of magnitude, it appears that dynamical horizon spacelike
expansions are probably the dominant mode in all but the most extreme 
situations, such as black hole and/or neutron star collisions. Numerical 
studies support this contention as in such extreme situations the 
occurrence of multiple horizons appears to be generic 
\cite{num3, badridavid} (though also relatively unstudied in a 
systematic way, since in most studies it is the exterior spacetime that 
is of  interest and so the interior of the outermost apparent horizon is 
excised and thrown away). By contrast for supermassive black holes, it is
likely that horizon jumps/TLMs are much more common. 
 
These examples also suggest other (possible) properties of MTTs. 
First, dynamical horizons appear to only originate either out of 
singularities (the density singularities of \ref{dustcollapse_sec}) or 
as part of a dynamical horizon/TLM pair (as in \ref{mtlm}). Equivalently, 
in all of our examples there is just one MTT associated with each black 
hole. This originates in a singularity and then weaves backwards and 
forwards in time, always expanding in area (relative to a foliation 
parameter that monotonically increases as one moves away from the 
singularity). This suggests that a similar result may be true away 
from spherical symmetry --- the multiple horizons/jumps seen in numerical 
studies of black hole collisions may actually all be part of a single 
MTT that weaves backwards and forwards in time. 

In our examples it is also true that TLMs and dynamical horizons always
occur in such a way that a causal signal originating from the MTT would never
be detectable by sufficiently far-away observers. This is consistent with
a gravitational confinement theorem due to Israel \cite{israel1,israel2}
which states that if the weak energy condition holds (as is the case in
all our examples), a trapped two-sphere can be extended to a spacelike
three-cylinder foliated by trapped two-spheres of \emph{constant area}. Assuming
reasonably regular spacetimes, this three-cylinder will act as a permanent
one-way membrane for causal effects. Even though an observer ``inside''
an MTT can escape that trapped region, he will not be able to 
send signals beyond the areal radius of the two-sphere on which the MTT
was first crossed. We note that Israel's confinement theorem is quite
general; it does not assume spherical symmetry or even asymptotic flatness.
Moreover, in appropriate asymptotically flat spacetimes, trapped surfaces must 
necessarily be contained within event horizons and so unable to send 
causal signals to null infinity \cite{hawkingellis72,wald}.
Our asymptotically flat examples are also consistent with the latter:
the outermost parts of MTTs are dynamical horizons which asymptote
to a null surface with some finite area.
Consequently, in physically realistic spacetimes, any TLMs that may be present 
will be hidden from observers far from a black hole, also in the absence of 
spherical symmetry.

In summary, even though the examples and (dust) calculations that we have 
seen in this paper are quite simple we believe that they are 
extremely useful in forming a correct intuition about the behaviour of 
MTTs during general black hole evolution. In particular they provide a 
convenient testing ground for ideas about these evolutions which is much 
simpler than full numerical simulations of black hole collisions and yet 
still significantly richer (and more realistic) than the heretofore 
studied analytical examples (Vaidya and Oppenheimer-Snyder). As such
they should be useful in, among other things,  
suggesting possible extensions of the recent mathematical 
investigations \cite{abhaygreg, ams05} of MTT properties.  


\section*{Acknowledgements:} The authors would like to thank 
Abhay Ashtekar, Chris Beetle, Steve Fairhurst, Greg Galloway, Sean Hayward,
Werner Israel, Badri Krishnan, Jos\'e Senovilla, the participants of Black Holes V and CCGRRA 11,  
and an anonymous referee who all made useful suggestions
and comments on this work during its development. I.~Booth was supported by NSERC. C.~Van 
Den Broeck was supported in part by the Eberly Research Fund of Penn State,
NSF grant PHY-00-90091, and the Edward M.~Frymoyer Honors Scholarship
Program. J.A.~Gonzalez  was supported by DFG grant ``SFB Transregio 7:
Gravitationswellenastronomie'' and by NSF grants PHY-02-18750 and 
PHY-02-44788.


\end{document}

%% file: SpacetimeCollapse.pstex_t
\begin{picture}(0,0)%
\includegraphics{SpacetimeCollapse.pstex}%
\end{picture}%
\setlength{\unitlength}{2960sp}%
\begingroup\makeatletter\ifx\SetFigFont\undefined%
\gdef\SetFigFont#1#2#3#4#5{%
  \reset@font\fontsize{#1}{#2pt}%
  \fontfamily{#3}\fontseries{#4}\fontshape{#5}%
  \selectfont}%
\fi\endgroup%
\begin{picture}(7060,4604)(2542,69)
\put(3909,1414){\rotatebox{45.0}{\makebox(0,0)[lb]{\smash{{\SetFigFont{11}{13.2}{\rmdefault}{\mddefault}{\updefault}{\color[rgb]{0,0,0}event horizon}%
}}}}}
\put(3944,2518){\makebox(0,0)[lb]{\smash{{\SetFigFont{11}{13.2}{\rmdefault}{\mddefault}{\updefault}{\color[rgb]{0,0,0}MTT}%
}}}}
\put(7653,2612){\makebox(0,0)[lb]{\smash{{\SetFigFont{11}{13.2}{\rmdefault}{\mddefault}{\updefault}{\color[rgb]{0,0,0}$\mathscr{I}^+$}%
}}}}
\put(4528,526){\makebox(0,0)[lb]{\smash{{\SetFigFont{11}{13.2}{\rmdefault}{\mddefault}{\updefault}{\color[rgb]{0,0,0}$\tau = \eta = 0$ (reflect through this axis)}%
}}}}
\put(4493,3985){\makebox(0,0)[lb]{\smash{{\SetFigFont{11}{13.2}{\rmdefault}{\mddefault}{\updefault}{\color[rgb]{0,0,0}$r=0$}%
}}}}
\put(2912,2125){\makebox(0,0)[lb]{\smash{{\SetFigFont{11}{13.2}{\rmdefault}{\mddefault}{\updefault}{\color[rgb]{0,0,0}$r=0$}%
}}}}
\end{picture}%

%% file: SpaceTimeAccretion.pstex_t
\begin{picture}(0,0)%
\includegraphics{SpaceTimeAccretion.pstex}%
\end{picture}%
\setlength{\unitlength}{3947sp}%
\begingroup\makeatletter\ifx\SetFigFont\undefined%
\gdef\SetFigFont#1#2#3#4#5{%
  \reset@font\fontsize{#1}{#2pt}%
  \fontfamily{#3}\fontseries{#4}\fontshape{#5}%
  \selectfont}%
\fi\endgroup%
\begin{picture}(5689,3566)(2745,874)
\put(4060,2033){\rotatebox{45.0}{\makebox(0,0)[lb]{\smash{{\SetFigFont{10}{12.0}{\rmdefault}{\mddefault}{\updefault}{\color[rgb]{0,0,0}event horizon }%
}}}}}
\put(3601,3989){\makebox(0,0)[lb]{\smash{{\SetFigFont{10}{12.0}{\rmdefault}{\mddefault}{\updefault}{\color[rgb]{0,0,0}$r=0$ spacelike singularity}%
}}}}
\put(2851,1664){\rotatebox{90.0}{\makebox(0,0)[lb]{\smash{{\SetFigFont{10}{12.0}{\rmdefault}{\mddefault}{\updefault}{\color[rgb]{0,0,0}continue with Schwarzschild}%
}}}}}
\put(3968,2494){\rotatebox{45.0}{\makebox(0,0)[lb]{\smash{{\SetFigFont{10}{12.0}{\rmdefault}{\mddefault}{\updefault}{\color[rgb]{0,0,0}MTT}%
}}}}}
\put(4077,1229){\makebox(0,0)[lb]{\smash{{\SetFigFont{10}{12.0}{\rmdefault}{\mddefault}{\updefault}{\color[rgb]{0,0,0}    $\tau = \eta = 0$ (reflect through this surface) }%
}}}}
\put(6970,2998){\makebox(0,0)[lb]{\smash{{\SetFigFont{10}{12.0}{\rmdefault}{\mddefault}{\updefault}{\color[rgb]{0,0,0}$\mathscr{I}^+$}%
}}}}
\end{picture}%